\documentclass{IEEEtaes}

\usepackage{color,array,amsthm}
\usepackage{graphicx}

\usepackage{cite}
\usepackage{amsmath,amssymb,amsfonts}

\usepackage{amsthm}
\usepackage{graphicx}
\usepackage{epstopdf}
\usepackage{xcolor}
\usepackage{multirow}
\usepackage{colortbl}

\usepackage{textcomp}
\usepackage{cite}
\usepackage{picinpar}
\usepackage{stfloats}
\usepackage{url}

\usepackage{import}
\usepackage{soul}
\usepackage{multirow}

\usepackage{pifont}
\usepackage{color}
\usepackage{alltt}

\usepackage{siunitx}
\usepackage{breakurl}
\usepackage{pbox}

\usepackage{algorithm,algpseudocode,float}
\usepackage{lipsum}
\usepackage{subfigure}
\usepackage{todonotes}

\usepackage{mathtools}
\usepackage{listings}                 
\usepackage{booktabs}
\usepackage{multirow}
\usepackage{balance}
\usepackage{float}
\usepackage{algorithm}
\usepackage{algpseudocode}

\usepackage[noabbrev,nameinlink,capitalise]{cleveref} 
\Crefformat{figure}{#2Fig.~#1#3}
\Crefmultiformat{figure}{Figs.~#2#1#3}{ and~#2#1#3}{, #2#1#3}{ and~#2#1#3}

\newcommand{\mb}[1]{\mathbf{#1}}



\jvol{XX}
\jnum{XX}
\jmonth{XXXXX}
\paper{1234567}
\pubyear{2020}
\doiinfo{TAES.2020.Doi Number}

\begin{document}

\title{A Visual Cooperative Localization Method for Airborne Magnetic Surveying Based on a Manifold Sensor Fusion Algorithm Using Lie Groups}

\author{Liang Liu}
\affil{ Shenyang Aerospace University, Shenyang, Liaoning, 110819, China} 

\author{Xiao Hu}
\affil{International Digital Economy Academy, Shenzhen 510085, China
} 

\author{Wei Jiang}
\member{Member, IEEE}
\affil{Beijing Jiaotong University, Beijing 100044, China}

\author{Guanglei Meng}
\affil{Shenyang Aerospace University, Shenyang, Liaoning, 110819, China}

\author{Zhujun Wang}
\affil{Shenyang Aerospace University, Shenyang, Liaoning, 110819, China}

\author{Taining Zhang}
\affil{Shenyang Aerospace University, Shenyang, Liaoning, 110819, China}

\receiveddate{Manuscript received XXXXX 00, 0000; revised XXXXX 00, 0000; accepted XXXXX 00, 0000.\\
This work was supported by the National Natural Science Foundation of China under Grant T2422002.
}

\corresp{
{\itshape (Corresponding author: Xiao Hu)}. 
}

\authoraddress{
Liang Liu, Guanglei Meng, Zhujun Wang, and Taining Zhang are with the School of Automation, Shenyang Aerospace University, Shenyang 110136, China (emails:  \href{liul062305@mail.nwpu.edu.cn}{liul062305@mail.nwpu.edu.cn}, \href{mengguanglei@yeah.net}{mengguanglei@yeah.net}, \href{wangzhujun22@163.com}{wangzhujun22@163.com}, \href{ztn6720@126.com}{ztn6720@126.com}).
Xiao Hu is an researcher at the Lower Airspace Economy Research Institute of International Digital Economy Academy, Shenzhen 510085, China (e-mail: \href{huxiao1@idea.edu.cn}{huxiao1@idea.edu.cn}).
Wei Jiang is with the State Key Laboratory of Rail Traffic Control and Safety and the Beijing Engineering Research Center of EMC and GNSS Technology for Rail Transportation, School of Electronic and Information Engineering, Beijing Jiaotong University, Beijing 100044, China (e-mail: \href{weijiang@bjtu.edu.cn}{weijiang@bjtu.edu.cn}).
}

\supplementary{Color versions of one or more of the figures in this article are available online at \href{http://ieeexplore.ieee.org}{http://ieeexplore.ieee.org}.}

\markboth{LIANG LIU ET AL.}{VISUAL COOPERATIVE LOCALIZATION IN UAV SURVEYING}
\maketitle

\begin{abstract}
Recent advancements in UAV technology have spurred interest in developing multi-UAV aerial surveying systems for use in confined environments where GNSS signals are blocked or jammed. This paper focuses airborne magnetic surveying scenarios. To obtain clean magnetic measurements reflecting the Earth's magnetic field, the magnetic sensor must be isolated from other electronic devices, creating a significant localization challenge. We propose a visual cooperative localization solution. The solution incorporates a visual processing module and   an improved manifold-based sensor fusion algorithm, delivering reliable and accurate positioning information. Real flight experiments validate the approach, demonstrating single-axis centimeter-level accuracy and decimeter-level overall 3D positioning accuracy.
\end{abstract}



\section{Introduction}
\label{sec:intro}
\begin{figure}
    \centering
    \includegraphics[width=0.25\textwidth]{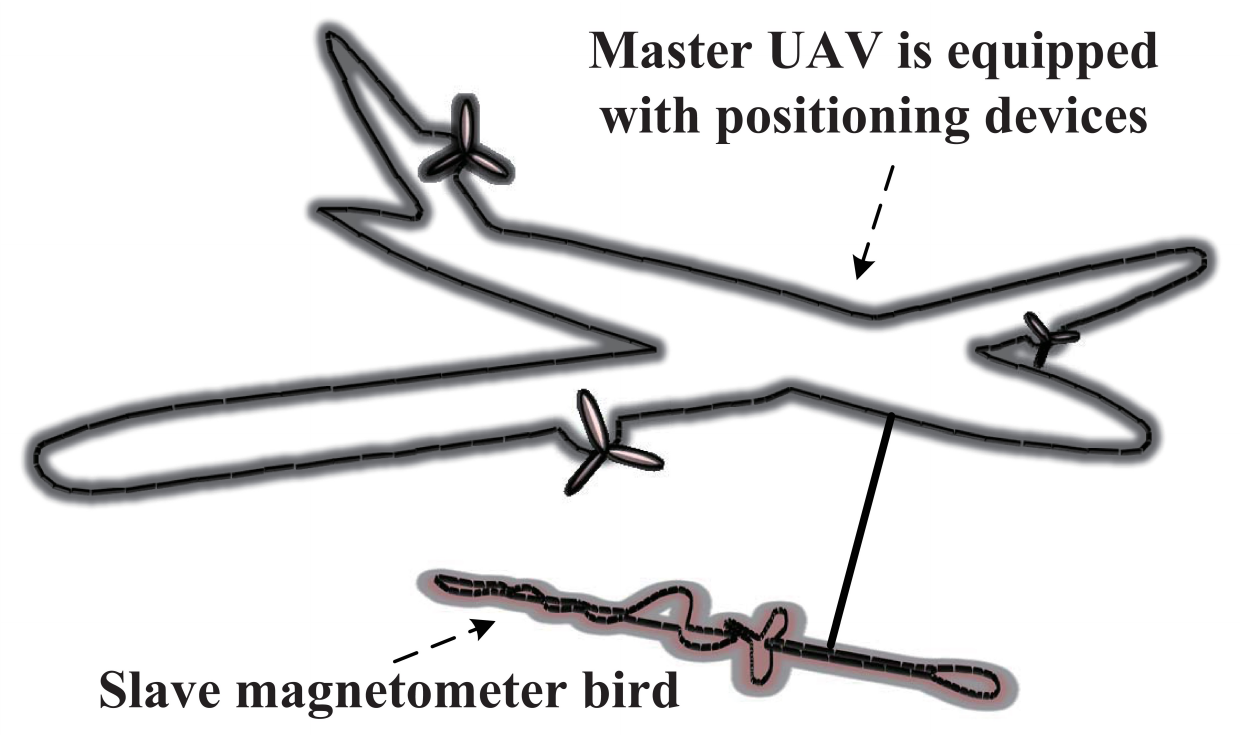}
    \caption[Conceptual illustration of the involved project.]{Conceptual illustration of the involved project.}
    \label{fig:figuav-qms}
\end{figure}

O{\scshape ver} the past two decades, Unmanned Aerial Vehicle (UAV) technology has garnered significant attention from both academia~\cite{kumar2012opportunities} and industry. With the rapid advancement of modern technologies, UAVs have seen widespread adoption in various civilian applications such as aerial photogrammetry, remote sensing~\cite{colomina2014unmanned}, and search and rescue missions~\cite{nagatani2013emergency}.
Despite these advancements, current UAV-based surveying technologies still rely heavily on Global Navigation Satellite Systems (GNSS) for precise positioning. This dependence poses serious challenges in GNSS-denied environments like underground tunnels, dense forests, or areas affected by interference, where autonomous UAV operation is limited or even unsafe. In response, research has shifted toward developing cooperative localization systems for multi-UAV setups~\cite{jospin2019photometric}. In these systems, UAVs equipped with GNSS can assist those without it by sharing position data, enabling safe and effective operation in GNSS-compromised zones. This cooperative localization approach becomes especially critical in magnetic surveying operations, where UAVs must operate in GNSS-denied environments, such as heavily mined regions or underground settings. UAV-based magnetic surveys require high-precision, noise-free magnetic data to accurately locate landmines or unexploded ordnance. Traditionally, such airborne magnetic surveys were performed using manned helicopters with suspended magnetometer sensors. 
These missions, though effective, are often expensive, labor-intensive, and hazardous. With advancements in UAV technology, the focus is shifting toward UAV-based solutions, offering safer, more flexible, and cost-efficient alternatives. A typical UAV-based magnetic surveying system, as depicted in~\cref{fig:figuav-qms}, comprises two main components: a master UAV and a slave UAV. The master UAV, equipped with GNSS and positioning devices, conducts manual or autonomous flights. The slave UAV, carrying only magnetic sensors, is suspended beneath the master UAV to collect magnetic data. 
By keeping the master and slave UAVs at a certain distance, the system ensures high-quality data with minimal magnetic interference. However, this setup presents two key challenges. First, precise localization of the slave UAV is critical to ensure magnetic isolation and minimal electronic interference while still geo-referencing the collected data. Second, the swing effect caused by the suspended slave UAV can impact the stability of the entire system.


In this paper, we address the challenge of accurately localizing the slave UAV while minimizing magnetic interference. We propose a cooperative localization solution that leverages a vision-based localization system, similar to previous work~\cite{bisgaard2010adaptive, hu2021toward, cucci2016accurate}. This approach involves estimating the position of the slave UAV relative to the master UAV using vision-based techniques. The local localization information is then fused with the global positioning data from the master UAV to determine the slave UAV’s global position. This solution minimizes additional electronic devices on the slave UAV (only an IMU is required on the slave UAV for attitude estimation), thus maintaining magnetic interference isolation, while still providing accurate global position estimates for the slave UAV. More specifically, the proposed solution incorporates a visual processing module designed for robust detection, tracking, and arc-based ellipse fiducial detection. This module is highly efficient and resilient to occlusions and varying perspectives, making it an ideal fit for the UAV-based magnetic surveying system, as depicted in~\cref{fig:figuav-qms}. Furthermore, we introduce a sensor fusion technique that employs an Unscented Kalman Filter (UKF) on manifolds, effectively fusing the global localization data from the master UAV with the local localization results from the slave UAV, ensuring precise and reliable position estimates.
The contributions of this paper are presented as follows:
\begin{enumerate}
\item We propose a vision-based cooperative localization system, providing a magnetic-interference-free positioning solution for the slave UAV with centimeter-level accuracy.
\item The visual processing module, in comparison to previous work \cite{bisgaard2010adaptive,hu2021vision}, demonstrates significantly higher robustness while maintaining efficiency and accuracy by incorporating visual tracking and arc-based ellipse detection.
\item For sensor fusion, we introduce a module using the UKF on manifolds. Unlike traditional approaches that assume alignment between the camera and UAV body frames, our method models the displacement between these frames and estimates a misalignment vector recursively. Additionally, we prove the observability of the proposed filter.
\end{enumerate}
Real flight experiments validate the effectiveness of the proposed solution. While developed for surveying, this approach is also applicable to slung-load state estimation and provides a foundation for similar algorithms in cargo transportation~\cite{bisgaard2007full,bisgaard2007vision}.

The paper's structure is outlined as follows: The related work and preliminaries are given in~\cref{sec:sf_rw} and \cref{sec:notations}, respectively. Section \ref{sec:proposed} delves into the methodology of the proposed approach. An experimental verification of the system is presented in Section \ref{sec:exp}. Concluding remarks are provided in Section \ref{sec:con}. Additionally, a comprehensive analysis of observability is carried out in Section \ref{sec:obs_type2}.

\section{Related Work}
\label{sec:sf_rw}
\subsection{Cooperative Localization \& Slung Load Estimator}
Cooperative localization methods in two-UAV systems has been investigated by~\cite{russell2019cooperative, cucci2016accurate,krajnik2014practical} for purposes including formation flight and photogrammetry.
~\cite{russell2019cooperative} proposed a bearing-only localization method via semi-definite relaxation and programming. While this method showed promise, it was time-consuming and validated only through simulations. Another approach by~\cite{jospin2019photometric} utilized active LEDs on a slave UAV, which were detected by cameras on a master UAV for cooperative localization. However, active targets like LEDs require continuous power, leading to challenges in power management and maintenance. In contrast, passive targets, such as planar-coded targets~\cite{cucci2016accurate}, address these issues by using geometric features like squares or circles embedded with codes to distinguish between multiple targets. These targets’ known dimensions, along with camera parameters, allow for the determination of relative positions from a single image. Circular markers, in particular, outperform square markers in detection range, making them a more accurate and reliable option for long-distance cooperative localization~\cite{krajnik2014practical}.

Estimating slung load states is also a domain that is relevant to this problem and has been widely studied, particularly to minimize swing effects during cargo transportation.~\cite{zhao2014robust} developed a vision-based system using the PnP algorithm, but with limitations. The pioneering work by~\cite{bisgaard2007vision, bisgaard2010adaptive} proposed a 3D pendulum model using two Euler angles with IMU data and a downward camera for updates, later extending this with an UKF to jointly estimate UAV and load states~\cite{bisgaard2007full}. This approach was validated with real flight data but had a drawback—it neglected the lever effects on IMUs, necessitating that the sensors be mounted near the UAV's center of mass (COM).


\subsection{Fiducial Detection}
Fiducial detection is a specific subset of general object detection, and techniques designed for general object detection can be applied to fiducial detection. However, general methods often prioritize universality over efficiency, making specialized fiducial detectors preferable in practice. Typically, fiducial detectors use classical image processing methods to identify specific geometric shapes and verify detected candidates through geometric constraints, image moments, or binary codes. For instance, ~\cite{krajnik2014practical} proposed a method for detecting circular patterns using adaptive thresholding, region growing, and verification of geometric constraints. 
~\cite{yang2013onboard} used adaptive thresholding and a neural network for circular fiducial detection.
Despite their real-time performance in embedded systems, these methods struggle with occlusions, motion blur, and require extensive parameter tuning for different fiducial marker configurations. For speeding up, ~\cite{krajnik2014practical} proposed to search for fiducials in the next frame based on their centers detected in the current frame. While these methods may be sufficient in some cases, they do not match the performance of advanced visual trackers.

\subsection{Ellipse Detection}
Ellipse detection is crucial in various applications due to the prevalence of ellipses in natural and human-made environments. However, noise and partial occlusions can significantly hinder the localization of ellipse edge pixels and split ellipse borders into discrete segments, making accurate and efficient detection challenging.
Ellipse detection methods are primarily categorized into Hough Transform (HT)-based~\cite{mclaughlin1998randomized} and arc-based approaches~\cite{fornaciari2014fast, puatruaucean2012parameterless, jia2017fast, lu2019arc, meng2020arc}. Arc-based methods have become the mainstream choice due to their superior accuracy and efficiency compared to HT-based methods.
Arc-based methods for ellipse detection focus on extracting arcs by chaining pixels based on constraints such as convexity~\cite{fornaciari2014fast}, curvature~\cite{meng2020arc}, and proximal connectivity~\cite{puatruaucean2012parameterless}. After extracting these arcs, they are grouped using convexity and positional constraints, followed by ellipse fitting and validation. These methods generally fall into three main categories, i.e. curvature based methods~\cite{ meng2020arc}, quadrant based methods~\cite{jia2017fast}, and parameterless methods~\cite{puatruaucean2012parameterless,lu2019arc} using the \textit{Contrario} technique.


\section{Preliminaries}
\label{sec:notations}
\subsection{Coordinate Systems}
\label{sec:coor_sys}
\begin{figure}
    \centering
    \includegraphics[width=0.45\linewidth]{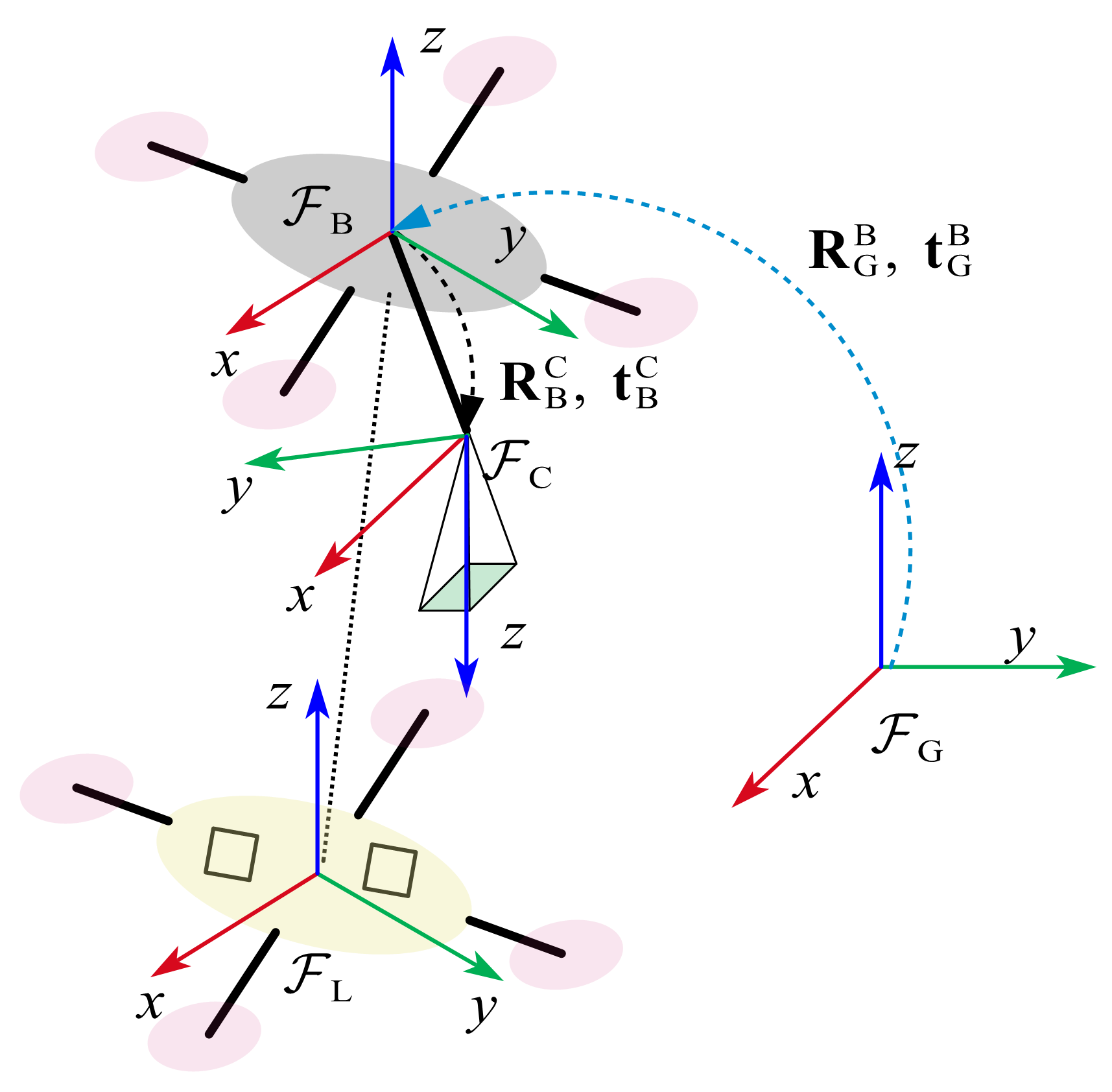}
    \includegraphics[width=0.35\linewidth]{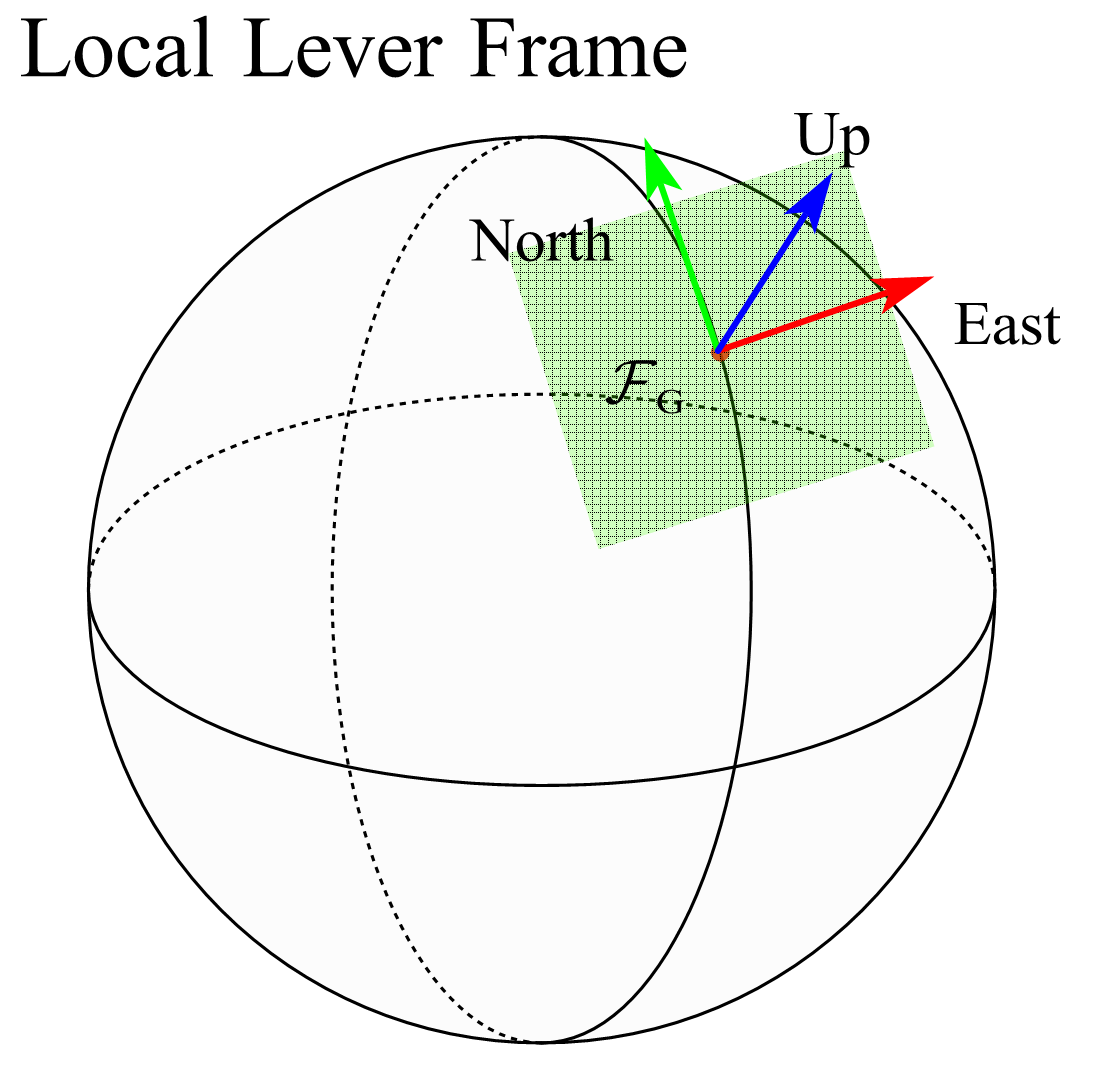}
    \caption[Illustration of the coordinate systems used in this work.]{Illustrations of the coordinate systems employed. The left part presents an overview of these coordinate systems, while the right part provides a more comprehensive depiction of the global coordinate frame. 
    }
    \label{fig:frames}
\end{figure}
\label{sec:cs}

This paper primarily utilizes the following coordinate systems, as depicted in Fig.~\ref{fig:frames}:
the \textbf{Global Coordinate System} ($\mathcal{F}_{\text{G}}$) is defined on a local level plane of a fixed point on the reference ellipsoid, oriented East ($x$), North ($y$), and Up ($z$), as shown in the right image of Fig.~\ref{fig:frames}. The \textbf{IMU Body Coordinate System} ($\mathcal{F}_{\text{B}}$) is fixed to the COM of a rigid body with axes pointing to the right, front, and up. The \textbf{Camera Coordinate System} ($\mathcal{F}_{\text{C}}$) has its origin at the camera center, with the $z$-axis along the optical axis, the $x$-axis pointing to the image's row direction, and the $y$-axis following the right-hand rule. The \textbf{Load Body Coordinate System} ($\mathcal{F}_{\text{L}}$) is fixed to the payload's COM, with axes corresponding to the right, front, and up directions of the body frame.





\subsection{Conics and Ellipses}
\label{sec:ellipsefitting}
The general equation of a conic in an affine plane can be expressed as follows:
\begin{equation}
\begin{aligned}
\mathbf{p}^{\top} \mathbf{C} \mathbf{p}=0
\end{aligned}
\label{eq:ellip_alg_dist}
\end{equation}
where $\mb{p} = [u, v, 1]^\top$ represents the point's homogeneous coordinates
and $\mathbf{C} \in \mathbb{R}^{3 \times 3}$ is the homogeneous matrix of a conic given as $\mathbf{C}=\left[A,\ B,\ D;\ B,\ C,\ E;\ D,\ E,\ F\right]$
where $A$ to $F$ are the coefficients of the conic. A conic is considered as an ellipse if the minor $\det(\mathbf{C}_{33}) = ac-b^2$ is greater than 0. The conic correspondence $\mb{C}^\prime$ of the conic $\mb{C}$ under 
a transformation $\mb{H}$ is given as
\begin{equation}
\mb{C}^\prime=\mathbf{H}^{-\top} \mathbf{C} \mathbf{H}^{-1}
\label{eq:h_conic}
\end{equation}

Ellipse fitting estimates the conic coefficients by minizing the point-to-conic distrance, either in the sense of algebraic or geomatric. The most well-known direct method was proposed by~\cite{fitzgibbon1999direct}, which minimizes the algebraic distance~\eqref{eq:ellip_alg_dist} under an additional constraint. The constrained optimization problem can be solved using the generalized eigenvalue decomposition. 

\subsection{Matrix Lie Group}
The rotation matrix and transformation matrix are examples of matrix Lie groups, which have the structure of a smooth manifold rather than a vector space~\cite{hauberg2013unscented}. The Special Orthogonal group $\mathbb{SO}(3)$ is the closed group of all $3 \times 3$ rotation matrices:


\begin{equation}
\label{eq:SO3definition}
\mathbb{SO}(3): \{{\mathbf{R}} \in \mathbb{R}^{3 \times 3}|\mathbf{R}\mathbf{R}^\top={\mathbf{I}},\ \det({\mathbf{R}})=1\}
\end{equation}
On any Lie group, the tangent space at the group identity has an associated Lie algebra $\mathfrak{g}$. The $\exp(\cdot)$ and $\log(\cdot)$ operations establish a local diffeomorphism between a neighborhood of $\mathbf{0}_{n \times n}$ in the tangent space and a local neighborhood of the identity on the manifold. The Lie algebra $\mathfrak{g}$ is linked to its vector space $\mathbb{R}^n$ by the mappings $({\cdot})^{\vee}: \mathfrak{g} \to \mathbb{R}^n$ and $({\cdot})^{\land}: \mathbb{R}^n \to \mathfrak{g}$. This property enables expressing differential calculus in vector space, which is necessary for optimization. The Lie algebra $\mathfrak{so}(3)$ is derived from the coefficient vector $\boldsymbol{\phi}\in \mathbb{R}^3$ as:
\begin{equation}
\begin{aligned}
\label{eq:so3alg}
\boldsymbol{\phi}^{\land}&=
\left[
\begin{matrix}
{\phi}_x \\ 
{\phi}_y \\ 
{\phi}_z
\end{matrix}
\right]^{\land}=\left[
\begin{matrix}
0 & -\phi_z & \phi_y \\
\phi_z & 0 & -\phi_x \\ 
-\phi_y & \phi_x & 0 \\
\end{matrix}
\right] \in \mathfrak{so}(3)
\end{aligned}
\end{equation}
On the $\mathbb{SO}(3)$, the exponential mapping exists as a closed-form formula:
\begin{equation}
    \exp(\boldsymbol{\phi}^{\wedge})=\mathbf{I}+\frac{\sin(||\boldsymbol{\phi}||)}{||\boldsymbol{\phi}||}\boldsymbol{\phi}^{\wedge}+\frac{1-\cos(||\boldsymbol{\phi}||)}{||\boldsymbol{\phi}||^2}\boldsymbol{\phi}^{\wedge}\boldsymbol{\phi}^{\wedge}
\end{equation}
where $||\boldsymbol{\phi}||$ is the standard Euclidean norm. Conversely, the logarithm mapping is given by
\begin{equation}
\log(\mathbf{R})=\frac{\theta \cdot (\mathbf{R}-\mathbf{R}^T)}{2\sin(\theta)}\ , \ \theta=\arccos(\frac{\text{tr}(\mathbf{R})-1}{2})
\label{eq:logso3}
\end{equation}
where $\text{tr}(\cdot)$ takes the trace of a matrix. 

\section{Proposed Method}
\label{sec:proposed}

\begin{figure}
    \centering
     \includegraphics[width=0.8\linewidth]{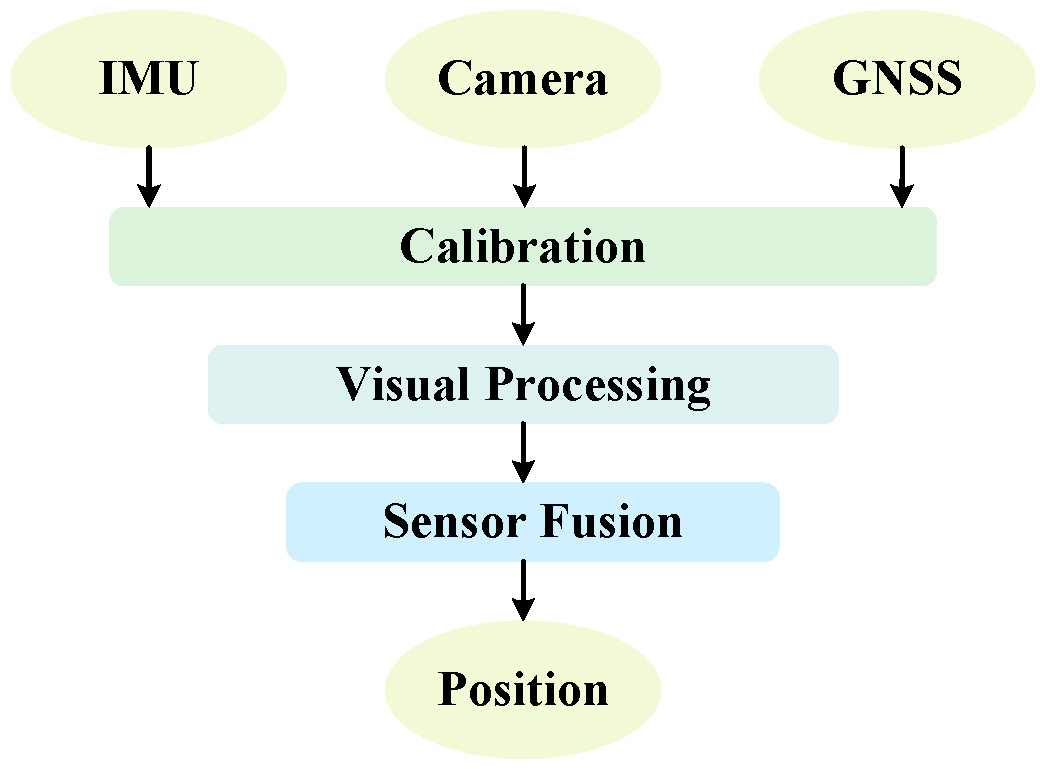}
    \caption{Flow chart of the vision-based cooperative localization system.}
    \label{fig:flowchart}
\end{figure}

The primary contributions of this paper lie in the visual processing and sensor fusion modules. This approach offers two main advantages: it is passive, ensuring excellent magnetic isolation, and it can provide the slave UAV's position even in GNSS-denied or featureless environments where direct positioning using GNSS or visual odometry may fail.
The overall flowchart of the vision-based positioning solution, shown in Fig.~\ref{fig:flowchart}, consists of three primary modules.
\begin{enumerate}
\item A temporal and spatial calibration module estimates parameters necessary for synchronizing and aligning measurements from GNSS, IMU, and the camera, allowing their use in the sensor fusion module. The calibration method from~\cite{hu2020multistate} is applied here.
\item A visual processing module employs detection, tracking, and arc-based ellipse detection techniques to efficiently detect elliptical fiducial markers from images.
\item A sensor fusion module extracts relative positions from fiducial markers and fuses them with global positions obtained from the GNSS/INS system using a UKF on manifold.
\end{enumerate}

\subsection{Visual Processing Module}
\begin{figure}[!ht]
    \centering
    \includegraphics[width=1.0\linewidth]{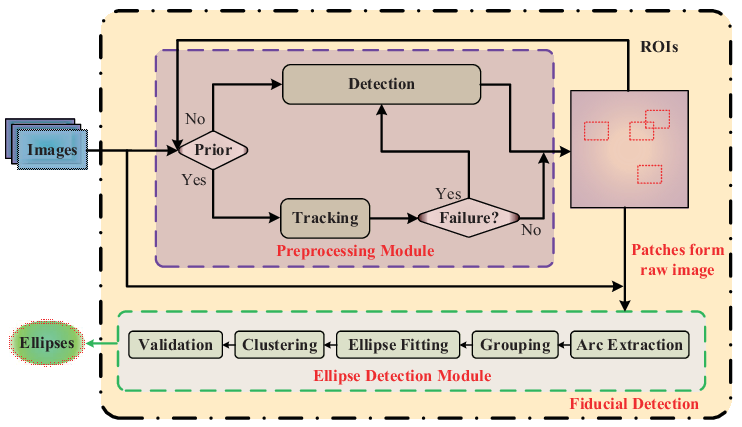}
    \caption{\textcolor{black}{Scheme chart of the fiducial detection module.}
    }
    \label{fig:scheme_fd}
\end{figure}

The objective of building the visual processing module is to efficiently detect the circular fiducial markers under variations of viewpoint, scale, illumination, and occlusion. Based on these factors, we developed a lightweight yet efficient fiducial detection algorithm that is robust under occlusions. Given that fiducials occupy only a small portion of the entire image (see examples in~\cref{sec:vis_pro}), it is inefficient to analyze the whole image. Conversely, downsampling may reduce accuracy due to detail loss. Therefore, this paper proposes a two-step fiducial detection process, as illustrated in Fig.\ref{fig:scheme_fd}. First, a preprocessing module identifies potential ROIs in the coarse image. Next, patches corresponding to these ROIs are cropped from the raw image. Finally, an ellipse detection module identifies valid ellipses within the patches.
Overall, this scheme improves efficiency while maintaining accuracy.


\subsubsection{Preprocessing Module}
For the preprocessing module, the Faster RCNN~\cite{ren2015faster} is used for fiducial object detection. When valid fiducial objects are identified, their associated trackers are initialized for visual tracking in the subsequent frames. We investigated four correlation filter-based visual trackers~\cite{mueller2017context,henriques2014high,danelljan2016discriminative,danelljan2017eco} and one color-based trackers~\cite{bertinetto2016staple} using our collected datasets. The results, detailed in~\cref{sec:tracking_exp}, led us to conclude that ECOHC~\cite{danelljan2017eco} is the best choice for prioritizing precision and robustness.
To avoid unconscious tracking loss or tracking wrong objects, we use the Peak-to-Sidelobe ratio~\cite{bolme2010visual}, which measures the strength of a correlation peak, to flag potential tracking failures. If tracking fails, the system switches back to detection to process the incoming image, preventing unconscious tracker drift.
 Experiments demonstrate that this framework achieves a rate of 40-90 Hz on images of 676×380 pixels on a laptop with an Intel i7 2.8 GHz CPU and an Nvidia A1000 GPU, sufficient for processing the 30 Hz video stream.

\subsubsection{Ellipse Detection Module}
The ellipse detection module utilizes an efficient arc-based ellipse detector, effectively addressing occlusion while maintaining efficiency. The process begins by extracting potential arcs in the image. These arcs are grouped if they likely belong to the same ellipse. For candidate arc groups, corresponding ellipses are fitted using the algebraic ellipse fitting algorithm proposed by~\cite{fitzgibbon1999direct}. On-the-fly pruning accelerates the grouping-fitting procedure. After generating a batch of candidate ellipses, clustering is performed to remove duplicates. Finally, all detected ellipses are validated based on several geometric indicators.
The proposed arc-based ellipse detection algorithm is based on\cite{lu2019arc,puatruaucean2012parameterless} with three major improvements: 1) arc segments selection using PCA instead of a batch of mannual defined thresholds; 2) iterative grouping, fitting, and pruning on sorted arc groups to efficiently reduce combinatorial trials; 3) clustering based on KD-tree. Due to the space limitation, we only introduce our contributions.

\begin{figure}[ht]
    \centering
    \subfigure[Illustration of using a PCA-based indicator to evaluate line segment straightness: as the radius increases (indicating greater straightness), the ratio also increases.]{
        \includegraphics[width=0.21\textwidth]{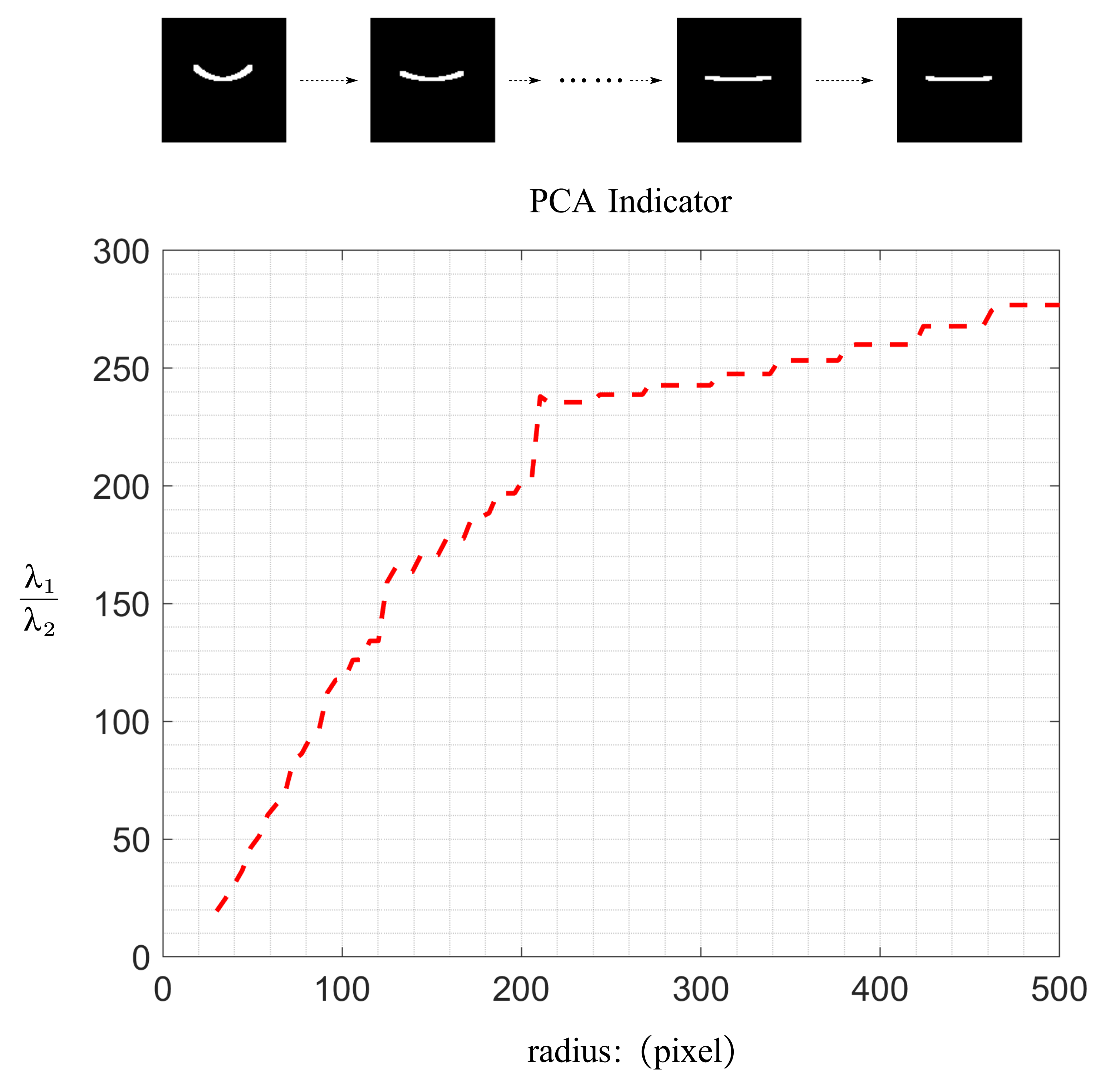}
        \label{fig:pca_lsd}
    }
    \hfill
    \subfigure[Illustration of span angle estimation:it is possible to estimate the ellipse center $\mb{c}$ using three points and their gradients. Once the center $\mb{c}$ is obtained, the span angle can be computed]{
        \includegraphics[width=0.21\textwidth]{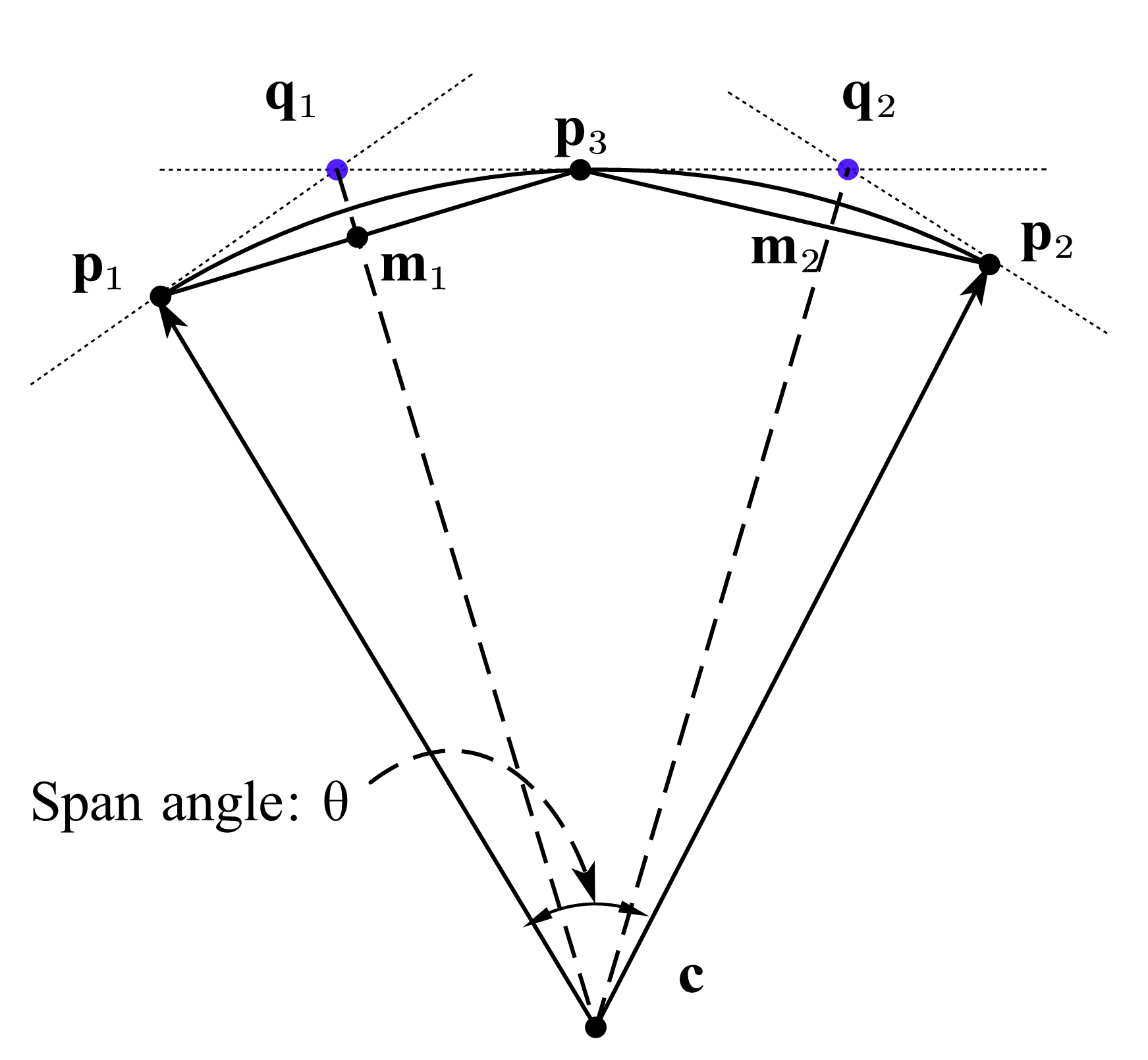}
        \label{fig:span}
    }
    \caption{Overall caption for both images}
    \label{fig:combined}
\end{figure}

\paragraph{PCA Based Arc Segment Extraction}
In literature, arcs are represented by either contours~\cite{fornaciari2014fast,jia2017fast} or connected line segments (LSs)~\cite{puatruaucean2012parameterless,lu2019arc,meng2020arc}. Due to potential issues with background noise in complex textures, we use the latter representation with the robust LSD detector~\cite{von2012lsd}. Straight LSs are unlikely to stem from arcs, so they can be pruned without further processing. Unlike~\cite{lu2019arc} that uses manually defined thresholds, we use PCA to distinguish straight LSs from non-straight ones by computing the ratio between the largest and smallest eigenvalues, serving as an indicator of straightness. A simulation was conducted to illustrate this method, as shown in~\cref{fig:pca_lsd}. Clearly, the straighter the arc, the larger the ratio. Thus, this indicator helps prune straight LSs while leaving other LSs for further processing. Importantly, computing this indicator does not increase computational complexity. The eigenvalues of the $2\times2$ covariance matrix can be solved analytically, and the covariance matrix is already computed in LSD, so no extra computation is required.

After extracting non-straight line segments (LSs), they are linked to form arcs based on proximity and curvature constraints to ensure smoothness as detailed in~\cite{puatruaucean2012parameterless}.  For each retrieved arc, its span angle can be estimated using the geometric method described by~\cite{yuen1989detecting}. The process, illustrated in~\cref{fig:span}, involves two points on the arc, and their gradients. By finding the midpoints ($\mb{m}_1$ and $\mb{m}_2$) and the intersection points ($\mb{q}_1$ and $\mb{q}_2$) of their tangents, the ellipse center can be approximated as lying on the line that passes through each midpoint and corresponding intersection point. Despite potential inaccuracies in the estimated center due to noise~\cite{fornaciari2014fast}, this approach provides a sufficient rough estimation of the span angle for ranking arcs, which is used in the following.

\begin{figure}
    \centering
    \includegraphics[width=0.5\linewidth]{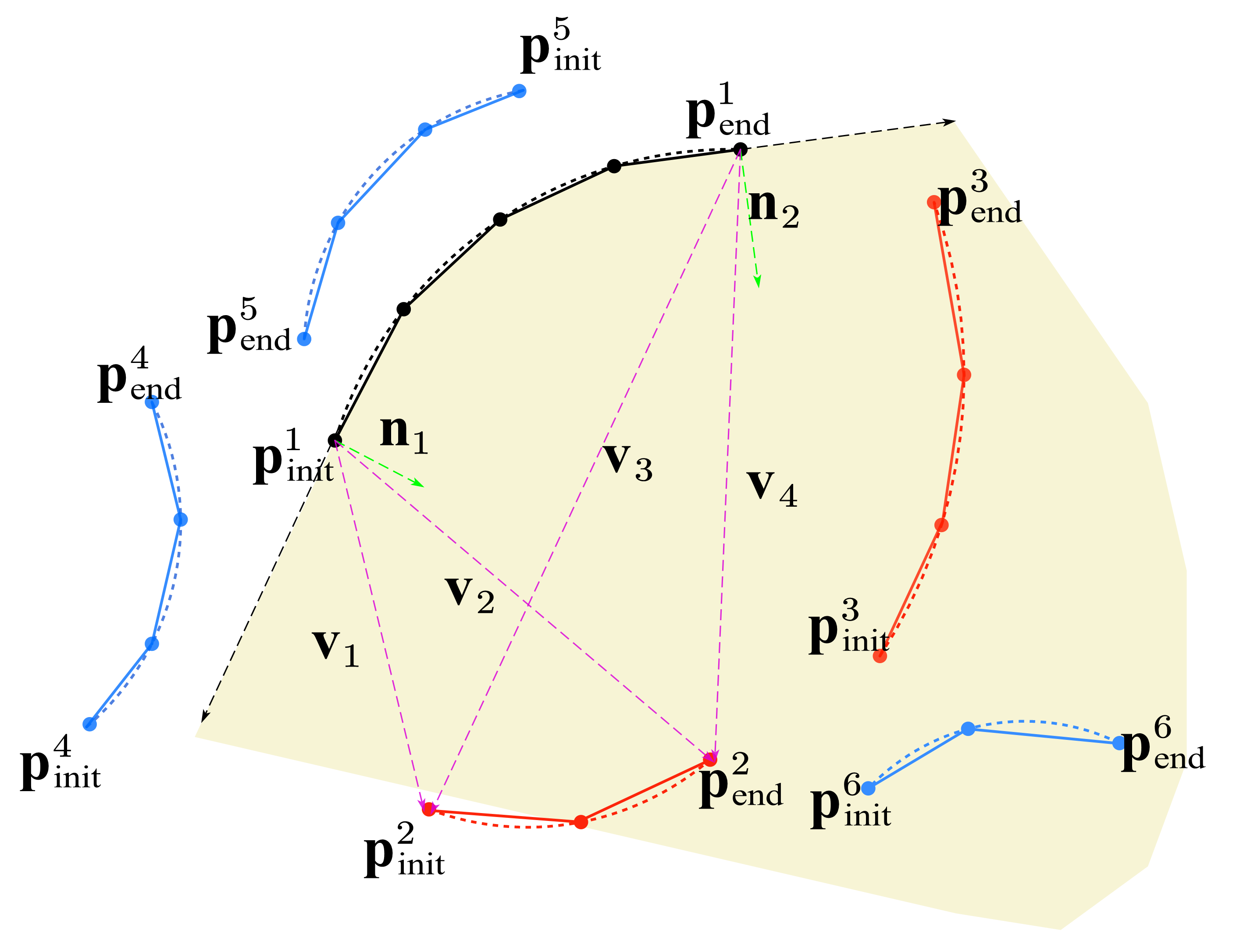}
    \caption[Illustration of region constraint used for arc grouping.]{Illustration of region constraint used for arc grouping: arc 1 (the query arc) can be grouped with arcs 2 and 3, but not with arcs 4 and 5, as they are outside the proper region (shaded area). Although arc 6 is in the region, arc 1 is not in the proper region formed by arc 6; thus, arc 6 cannot be grouped with arc 1.}
    \label{fig:convexity}
\end{figure}

\paragraph{Grouping and On-the-fly Pruning}
Fitting ellipses with incomplete data, such as small arcs, often leads to poor estimates. Therefore, associating arcs that may belong to the same ellipse is crucial. Strategies based on region analysis~\cite{meng2020arc, lu2019arc} are preferred due to their clear geometric interpretation, and this approach is employed here for arc grouping. As illustrated in~\cref{fig:convexity}, arc 1 can be grouped with arc 2 and arc 3, but not with arc 4 and arc 5, which are outside the proper region (shaded area). This grouping requires checking the dot products of the corresponding vectors, as follows:
\begin{equation}
    \mb{n}_1\mb{v}_1 \geq 0,\ \mb{n}_1\mb{v}_2 \geq 0,\ \mb{n}_2\mb{v}_3 \geq 0,\ \mb{n}_2\mb{v}_4 \geq 0
\end{equation}
Here, $\mb{v}_i$ denotes a vector composed of the initial point of one arc to the initial or endpoint of another arc, and $\mb{n}_i$ represents a normal vector of a LS. A one-way check cannot rule out exceptions, such as arc 6 shown in \cref{fig:convexity}. However, a mutual region constraint check can rule out such exceptions. This constraint is a necessary condition, meaning arcs meeting this criterion are not necessarily from the same ellipse. Instead of grouping arcs in an arbitrary order, we first rank arcs by their span angles and start with the arc having the maximum span angle. For each group established, we check the sum of the span angles. If the total span angle exceeds a threshold, ellipse fitting is applied to this arc group immediately. After fitting and fast validation, the ellipse is used to prune other arcs that fail to associate with this arc group but lie on the same ellipse. 


The on-the-fly pruning process prunes arcs on the same ellipse but not included in the initial arc group for fitting. Since ellipses from incomplete data are unreliable, pruning is only applied when a salient ellipse is detected. Salience is determined by the total span angle of the arcs used for fitting, recalculated using the fitted ellipse center. If the span angle exceeds a threshold (e.g., $135^\circ$), the ellipse is considered salient. To prune ungrouped arcs, the mean algebraic distance to the identified ellipse is computed. Arcs with a mean algebraic distance below a certain criterion are added to the group and removed from the list of arcs awaiting grouping. If additional arcs are added, we refit an ellipse. This iterative grouping-fitting-pruning technique reduces the number of trials needed to find all ellipses.

\paragraph{Clustering}
To eliminate duplicate detections, a clustering post-processing step is applied. An ellipse is represented in its parametric form, making it a five-dimensional point. However, this five-dimensional space is not a vector space due to the orientation angle. Unlike~\cite{lu2019arc}, which uses a hierarchical method for clustering in divided spaces, we propose a more efficient approach: clustering first in a 4D space (ellipse center, semi-major, and semi-minor axes) and then in a 1D space (orientation).
We utilize the mean shift algorithm~\cite{cheng1995mean} for clustering. Given that the mean shift algorithm requires repeated queries for the nearest neighbors and neighbors within a window, we use a KDTree to accelerate this process.

\subsubsection{Final Output: Conic Matrix}
For each validated ellipse, its conic matrix is computed in the local coordinate system of the patch. To use this matrix for positioning in the original image, it must be transformed accordingly. Given that points on the local patch correspond to points on the original image via a translation, let $\mathbf{u}_2=[x_2, y_2, 1]^\top$ denote a point on the local patch and $\mathbf{u}_1=[x_1, y_1, 1]^\top$ the corresponding point on the original image. Their relationship can be established using a transformation matrix $\mb{T}$ as follows:
$$
\mathbf{u}_2 = \mb{T}\mathbf{u}_1,\quad \text{where } \mb{T}=\left[
\begin{array}{ccc}
    1 &  0 & -\Delta_x \\
    0 & 1 & -\Delta_y \\
    0 & 0 & 1
\end{array}
\right]
$$
where $\Delta_x, \Delta_y$ are the offsets that make $x_2 = x_1-\Delta_x,y_2 = y_1-\Delta_y$. The corresponding conic matrix $\mb{C}_2$ on the original image can be computed from the conic matrix $\mb{C}_1$ on the local patch by~\eqref{eq:h_conic}, i.e. $\mb{C}_2 = \mb{T}^\top \mb{C}_1 \mb{T}$.
The conic matrix $\mb{C}_2$ is the final output of the fiducial detection module and will be used in~\cref{sec:sensorfusion} for positioning.

\subsection{Sensor Fusion Module}
\label{sec:sensorfusion}
In the sensor fusion module, we developed a recursive filter on a manifold. Compared to our previous work~\cite{hu2021toward}, this improved filter also incorporates data from the IMU on the payload, allowing the use of a general motion model as the process model. This eliminates the need to assume that the fiducial center coincides with the COM. Theoretically, the proposed filter provides higher quality results by exploiting more information and building upon the UKF on manifold. The design and development of this filter are detailed in the following sections.


\subsubsection{Resolving Relative Position}
\label{sec:RelativePosition}
The position of the circle center in relation to the camera coordinate frame can be determined from the extracted conic matrix $\mb{C}$~\cite{kanatani19933d}, involves using the projected ellipse and its known radius. The procedures are briefly outlined. By applying eigenvalue decomposition on $\mathbf{C}$, we have:
\begin{equation}
\mathbf{C}=\mathbf{V} \mathbf{\Lambda} \mathbf{V}^{\mathrm{T}}, \mathbf{\Lambda}=\left[\begin{array}{ccc}
\lambda_{1} & 0 & 0 \\
0 & \lambda_{2} & 0 \\
0 & 0 & \lambda_{3}
\end{array}\right], \mathbf{V}=\left[\begin{array}{lll}
\mathbf{v}_{1} & \mathbf{v}_{2} & \mathbf{v}_{3}
\end{array}\right]
\end{equation}
where $\mathbf{\Lambda}$ is the diagonal matrix formed by the eigenvalues, and $\mathbf{V}$ is a unitary matrix whose column vectors correspond to the eigenvectors. \cite{kanatani19933d} proved that $\mathbf{C}$ is a real conic if and only if it has a signature of $\{2,1\}$. Without loss
of generality, by assuming that $\lambda_{1} \geq \lambda_{2}>0>\lambda_{3}$, the position of the circle center and the normal of the circular plane with respect to the camera coordinate frame can be computed as 
\begin{equation}
\begin{aligned}
\mathbf{p}&=\frac{s_3\mathbf{r}}{\sqrt{-\lambda_{1} \lambda_{3}}}\left(s_{1} \lambda_{3} \sqrt{\frac{\lambda_{1}-\lambda_{2}}{\lambda_{1}-\lambda_{3}}} \mathbf{v}_{1}+s_{2} \lambda_{1} \sqrt{\frac{\lambda_{2}-\lambda_{3}}{\lambda_{1}-\lambda_{3}}} \mathbf{v}_{3}\right) \\
\mathbf{n}&=\left(s_{1} \sqrt{\frac{\lambda_{1}-\lambda_{2}}{\lambda_{1}-\lambda_{3}}} \mathbf{v}_{1}+s_{2} \sqrt{\frac{\lambda_{2}-\lambda_{3}}{\lambda_{1}-\lambda_{3}}} \mathbf{v}_{3}\right)
\end{aligned}
\end{equation}
where $r$ represents the radius of the circle, $s_1, s_2, s_3$ denote three undetermined signs. With these parameters, applying the chirality constraint, which ensures that the object lies in front of the camera, and assuming the normal points towards the camera optical center, allows us to resolve two out of the three signs:
\begin{equation}
\begin{array}{l}
\mathbf{n}^\top \mb{e}_3<0 \\
\mathbf{p}^\top \mb{e}_3>0, \text{ where } \mb{e}_3=[0,0,1]^\top
\end{array}
\end{equation}
The remaining sign can be determined with the help of a second fiducial marker. 
Given that each fiducial marker can have two possible normals, we resolve the unknown sign which results in the minimal included angle between the two normals:
\begin{equation}
\begin{array}{l}
i^*, j^*: \underset{i,j}{\text{ argmax }} {\mathbf{n}_{1i}}^\top \mb{n}_{2j}
\end{array}
\end{equation}
where $\mb{n}_{1i}, i=1,2$ and $\mb{n}_{2j}, j=1,2$ are normals from the first and second fiducial marker, respectively.
\subsubsection{UKF on Lie Groups}
The backbone of the developed filter is the UKF on Lie Groups, which is introduced in the following and summarized in~\cref{alg:ukflg}.
The concentrated Gaussian distribution defines a locally approximated Gaussian distribution on a Lie group with the help of a corresponding Gaussian distribution defined in a tangent space~\cite{barfoot2017state}
, which can be written as
\begin{equation}
\boldsymbol{\chi}=\boldsymbol{\bar{\chi}} \exp (\boldsymbol{\xi}^\land), \boldsymbol{\xi} \sim \mathcal{N}(\mathbf{0}, \mathbf{P})
\label{eq:cgd}
\end{equation}
where $\boldsymbol{\xi}$ is a vector in the tangent space following the Gaussian distribution $ \mathcal{N}(\mathbf{0}, \mathbf{P})$, $\boldsymbol{\chi}$ is the corresponding random variable on the Lie group that follows the concentrated Gaussian distribution $\mathcal{N}(\boldsymbol{\bar{\chi}}, \mathbf{P})$. An alternative definition using right multiplication can be derived~\cite{barfoot2017state}. 
With the help of the concentrated Gaussian distribution, the UKF on Lie groups can be worked out subsequently. If the state consists of both Lie group variables and normal vector variables, the UKF on Lie groups can be generalized by treating group variables and vector variables separately according to their corresponding operations. We summarize the generalized UKF algorithm on Lie groups using left multiplication in \cref{alg:ukflg}. The $\oplus$ operator in~\cref{alg:ukflg} is defined as follows: $\oplus=\exp(\cdot^\land)$ for group variables and $\oplus=(\cdot)+(\cdot)$ for vector variables. Similarly, $\ominus=\log(\cdot^\vee)$ for group variables and $\ominus=(\cdot)-(\cdot)$ for vector variables.

\begin{algorithm}[!htp]
\caption{Unscented Kalman Filter on Lie Groups} 
\label{alg:ukflg}
\begin{algorithmic}[1]
\State \textbf{Input: } $\bar{\boldsymbol{\chi}}_{k-1}$, $\mathbf{P}_{k-1}$, f, $\mb{u}_k$,  $\mathbf{Q}_{k}$, h, $\mathbf{y}_{k}$, $\mathbf{R}_{k}$.
\State \textbf{Output: } $\bar{\boldsymbol{\chi}}_k$, $\mathbf{P}_{k}$.
\Function{Propagation}{$\bar{\boldsymbol{\chi}}_{k-1}$, $\mathbf{P}_{k-1}$, f, $\mb{u}_k$,  $\mathbf{Q}_{k}$}
 \State $\mathbf{P}^{\text {aug }}=\operatorname{diag}(\mathbf{P}_{k-1}, \mathbf{Q}_{k})$ 
\State Initialize sigma weights $W_i^{m},\quad W_i^{c}$ according to~\cite{julier1997new}.
\State Initialize sigma increments analogous to~\cite{julier1997new}:
$$\Delta \boldsymbol{\chi}_0=\mb{0}, \Delta \boldsymbol{\chi}_i= \pm \left(\sqrt{(N+\lambda) \mathbf{P}_{\text {aug }}}\right)_{i},\ i=1, \ldots, 2N$$
\State Formulate sigma points by $\boldsymbol{\chi}^{s_i}_{k-1}=\bar{\boldsymbol{\chi}}_{k-1}\oplus \Delta \boldsymbol{\chi}_i$.
\State Propagate sigma points through the process model $\boldsymbol{\chi}^{s_i}_{k}=f(\boldsymbol{\chi}^{s_i}_{k-1},\mb{u}_k)$ and the mean and covariance are updated as $
\bar{\boldsymbol{\chi}}_{k|k-1}=\boldsymbol{\chi}^{s_0}_{k},\ 
\mathbf{P}_{k|k-1} = \sum_{i=0}^{2 N} W_{i}^{c}\boldsymbol{\epsilon}_{\boldsymbol{\chi}}^i{\boldsymbol{\epsilon}_{\boldsymbol{\chi}}^i}^{\top},
$
where $\boldsymbol{\epsilon}_{\boldsymbol{\chi}}^i=\boldsymbol{\chi}^{s_i}_{k}\ominus \bar{\boldsymbol{\chi}}_{k|k-1}$.
\EndFunction
\Function{Update}{$\bar{\boldsymbol{\chi}}_{k|k-1}$, $\mathbf{P}_{k|k-1}$, h, $\mathbf{y}_{k}$, $\mathbf{R}_{k}$}
\State $\mathbf{P}^{\text {aug }}=\operatorname{diag}(\mathbf{P}_{k|k-1}, \mathbf{R}_{k})$ 
\State Initialize sigma weights $W_i^{m},\quad W_i^{c}$ according to~\cite{julier1997new}.
\State Initialize sigma increments analogous to~\cite{julier1997new}: 
$$\Delta \boldsymbol{\chi}_0=\mb{0}, \Delta \boldsymbol{\chi}_i= \pm \left(\sqrt{(N+\lambda) \mathbf{P}_{\text {aug }}}\right)_{i},\ i=1, \ldots, 2N$$
\State Formulate sigma points by $\boldsymbol{\chi}^{s_i}_{k}=\bar{\boldsymbol{\chi}}_{k|k-1}\oplus \Delta \boldsymbol{\chi}_i$.
\State Propagate sigma points through the process model  $\mathbf{y}^{s_i}_{k}=h(\boldsymbol{\chi}^{s_i}_{k})$.
\State Compute the output mean and covariance matrices as $
\bar{\mathbf{y}}=\underset{\bar{\mathbf{y}}}{\text{argmin }}\sum_{i=0}^{2N} W^{m}_{i} (\mathbf{y}^{s_i}_{k} \ominus \bar{\mathbf{y}}),\ 
\mathbf{P}_{\mathbf{y y}}=\sum_{i=0}^{2N} W^{c}_{i}\boldsymbol{\epsilon}_{\mathbf{y}}^i{\boldsymbol{\epsilon}_{\mathbf{y}}^i}^{\top},\ \mathbf{P}_{\boldsymbol{\chi}\mathbf{y}}=\sum_{i=0}^{2N} W^{c}_{i}
\boldsymbol{\epsilon}_{\boldsymbol{\chi}}^i
{\boldsymbol{\epsilon}_{\mathbf{y}}^i}^\top, 
$
where $\boldsymbol{\epsilon}_{\mathbf{y}}^i=\mathbf{y}^{s_i}_{k}\ominus\overline{\mathbf{y}}$.
\State Compute the innovation: $\boldsymbol{\xi} =\mathbf{P}_{\boldsymbol{\chi}\mathbf{y}} \mathbf{P}_{\mathbf{y y}}^{-1}(\mathbf{y}_k \ominus \bar{\mathbf{y}})$.
\State State and covariance update:
$
\bar{\boldsymbol{\chi}}_k=\bar{\boldsymbol{\chi}}_{k|k-1} \oplus (\boldsymbol{\xi}),\quad \mathbf{P}_k=\mathbf{P}_{k|k-1}-\mathbf{P}_{\boldsymbol{\chi}\mathbf{y}}\left(\mathbf{P}_{\boldsymbol{\chi}\mathbf{y}} \mathbf{P}_{\mathbf{y y}}^{-1}\right)^{\top}
$.
\EndFunction
\end{algorithmic}
\end{algorithm}

\subsection{Proposed Filter}
\label{sec:filter}
The state vector of the proposed filter is specified as
\begin{equation}
    \mathbf{x}=\{\mb{R}^{\text{G}}_{\text{L}},\ [\mb{p}^{\text{G}}_{\text{L}},\ \mb{v}^{\text{G}}_{\text{L}},\ l,\ \Delta \mb{t}^{\text{C}}_{\text{B}},\ \mb{t}^{\text{L}}_{\text{F}_i}]\}, \text{ where } i=1,\cdots,m
\end{equation}
where $\mb{R}^{\text{G}}_{\text{L}}$, $\mb{p}^{\text{G}}_{\text{L}}$, $\mb{R}^{\text{G}}_{\text{L}}$ denotes the rotation matrix, position, and velocity of the load body coordinate frame with respect to the global coordinate frame, respectively, l is the rope length, $\Delta \mb{t}^{\text{C}}_{\text{B}}$ is the misalignment term of the relative translation, and $\mb{t}^{\text{L}}_{\text{F}_i}$ represents the relative translation from the center of the $i^{\text{th}}$ fiducial marker to the origin of the load body coordinate frame. L and F denotes the load and fiducial, respectively. $m$ is the total number of fiducials.

\subsubsection{Process Model}
The process model describes the general motion of the system, governed by IMU measurements mounted on the load:
\begin{equation}
    \begin{aligned}
        \dot{\mb{R}}^{\text{G}}_{\text{L}}&=\mb{R}^{\text{G}}_{\text{L}}(\boldsymbol{w}+\mb{n}_w)^\land\\
        \dot{\mb{p}}^{\text{G}}_{\text{L}}&=\mb{v}^{\text{G}}_{\text{L}}\\
        \dot{\mb{v}}^{\text{G}}_{\text{L}}&=\mb{R}^{\text{G}}_{\text{L}}(\mb{a}+\mb{n}_a)-\mathbf{g}\\
        \dot{l}&=n_1,\quad \Delta \dot{\mb{t}}^{\text{C}}_{\text{B}}=\mb{n}_2\quad \dot{\mb{t}}^{\text{L}}_{\text{F}_i}=\mb{n}_3
    \end{aligned}
\end{equation}
where $\boldsymbol{w}$ and $\mb{a}$ denote the angular velocity and acceleration measured by the IMU, respectively, $\mb{n}_w$, $\mb{n}_a$, $n_1$, $\mb{n}_2$, and $\mb{n}_3$ represent corresponding white Gaussian noise. $l,\ \Delta \mb{t}^{\text{C}}_{\text{B}},\ \mb{t}^{\text{F}}_{\text{L}_i}$ are modeled as slow time-varying parameters. 

\subsubsection{Measurement Model}
The measurement model consists of three distinct measurements utilized in the proposed filter, each capturing specific aspects of the system state for accurate estimation.
Rotation Measurement ($\mb{y}_{\mb{R}}$): This measurement captures the relative orientation between the reference orientation matrix $\bar{\mb{R}}^{\text{G}}_{\text{L}}$ and the load body orientation $\mb{R}^{\text{G}}_{\text{L}}$. It helps in accurately tracking the orientation of the load body relative to a known reference.
Position Measurement ($\mb{y}_{\mb{p}_{i}}$): For each fiducial marker ($i$) on the load, this measurement captures the position of the marker in the global coordinate system. The term $\mb{t}^{\text{C}}_{\text{B}}$ represents the translation from the UAV's center to the load body's center, and $\Delta \mb{t}^{\text{C}}_{\text{B}}$ denotes the misalignment of the relative translation that can be estimated and compensated online. This measurement facilitates accurate localization of the load and provides information about its position with respect to the UAV and the global coordinate frame.
Virtual Length Measurement ($\mb{y}_{l}$): this measurement uses the equality constraint $|\mb{p}^{\text{G}}_{\text{L}} - \mathbf{p}^{\text{G}}_{\text{B}}| = l$ as a virtual measurement to update the formation distance.
\begin{equation}
    \begin{aligned}
    \mb{y}_{\mb{R}} &= \log({\bar{\mb{R}}^{\text{L}}_{\text{G}}} \mb{R}^{\text{G}}_{\text{L}})^\vee \\
    \mb{y}_{\mb{p}_{i}} &= \mb{R}^{\text{C}}_{\text{G}}(\mb{p}^{\text{G}}_{\text{L}}+\mb{R}^{\text{G}}_{\text{L}}\mb{t}^{\text{L}}_{\text{F}_i}-\mb{p}^{\text{G}}_{\text{B}})+\mb{t}^{\text{C}}_{\text{B}}+\Delta \mb{t}^{\text{C}}_{\text{B}}\\
     \mb{y}_{l} &= \|\mb{p}^{\text{G}}_{\text{L}}-\mb{p}^{\text{G}}_{\text{B}}\|-l
    \end{aligned}
\end{equation}
Due to the asynchronous nature of these measurements, the proposed filter updates the state estimation sequentially. Whenever a subsequent measurement becomes available, the corresponding measurement update is applied to the filter, enabling continuous and accurate tracking of the system state.

\subsubsection{Observability}
Ensuring system observability is crucial for the proposed filter. A thorough analysis, detailed in Appendix~\cref{sec:obs_type2}, involves constructing the observation space using Lie derivatives to assess observability. The analysis confirms that the proposed filter is observable, meaning available measurements sufficiently estimate the entire system state, including hidden states like the misalignment term and fiducial marker positions.

\section{Experiments}
\label{sec:exp}
\begin{figure}
    \centering
    \includegraphics[width=1.0\linewidth]{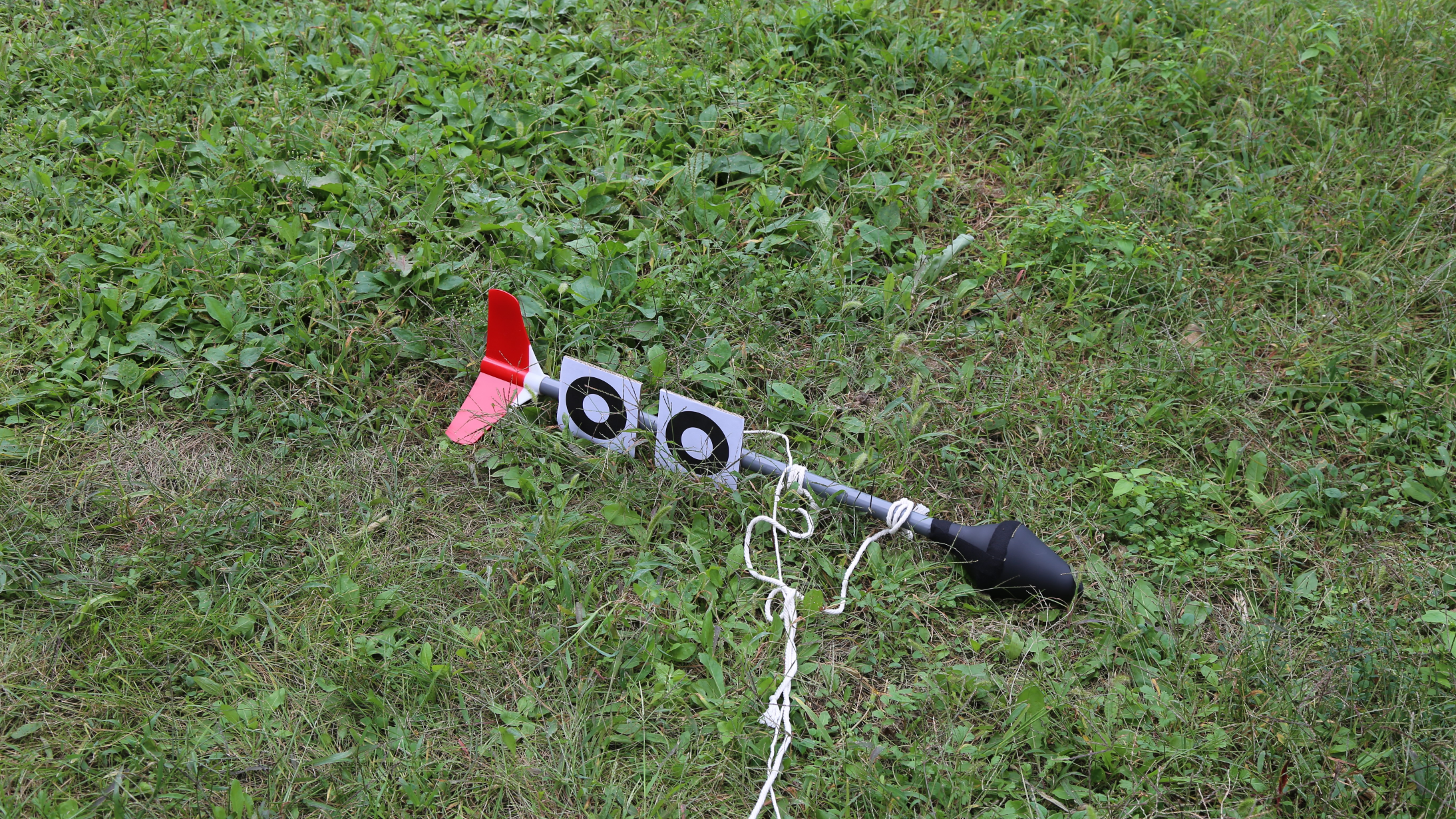}
    \caption[System setup of the field flight tests.]{The system setup for the field flight tests includes a prototype magnetometer with two fiducial markers mounted on it.}
    \label{fig:exp_setup}
\end{figure}
To evaluate the proposed solution, we conducted a field experiment using a DJI Matrice 600 UAV, which carried a slave magnetometer bird suspended 7 meters below, as shown in~\cref{fig:exp_setup}. This setup was used to simulate a magnetic surveying scenario. During the experiment, three flights were performed, each with two different fiducial marker configurations. The sensors deployed in the experiment included:
\begin{itemize}
    \item On the UAV: Novatel SPAN GNSS/INS system (ADIS16488 IMU) with Tallysman TW3972 antenna and GoPro Hero12 camera (30 FPS, 2704 $\times$ 1520 resolution).
    \item On the magnetometer assembly: Two fiducial markers and another Novatel SPAN GNSS/INS system (ADIS 16488) for reference.
\end{itemize}
The UAV was operated manually during the experiment, with the flight path designed to replicate a practical surveying trajectory. The GNSS/INS data was processed with Novatel's Inertial Explorer software, using a local base station for relative carrier-phase solutions. An overview of fiducial marker parameters is provided in~\cref{tb:exp_setup}.

\begin{table}[!htb]
	\centering
	\caption{Fiducial Marker Specifications Used for Flight Tests.}
	\begin{tabular}{c||c|c|c}
		\hline
		\textbf{Parameter} & \multicolumn{3}{c}{\textbf{Value (cm)}} \\ \cline{1-4}
        & \textbf{1st} & \textbf{2nd} & \textbf{3rd}   \\ \cline{1-4}
		Diameter (Outer Circle)  & 21.96 & 26.84 & 26.84  \\ \hline
		Diameter (Inner Circle)  & 9 & 11 & 11 \\ \hline
	\end{tabular}
	\label{tb:exp_setup}
\end{table}

\subsection{Experiments on visual procssing module}
\label{sec:vis_pro}
\subsubsection{Tracking Evaluation Results}
\label{sec:tracking_exp}
To evaluate the trackers' performance, we tested them on collected datasets, each containing a raw video sequence of 2 to 2.5 minutes. Trackers were initialized using a detected bounding box, and no further detection was performed after initialization, even if a tracker failed. We report the first tracking failure time for each tracker to assess endurance and robustness, along with the FPS as a measure of efficiency. Quantitative results are shown in~\cref{tb:visual_tracking_own}. Our observations indicate that KCF is the fastest tracker, while ECOHC is the most robust, though the slowest.

\begin{table}[!ht]
	\centering
	\caption{{Evaluation results of concerned visual trackers on collected datasets.}}
	\begin{tabular}{c|c|c|c|c}
		\hline
		& \multicolumn{3}{c}{\textbf{Failure time: (s)}}  & \textbf{Mean FPS} \\ \cline{2-5}
        \textbf{Sequence Number} & {1st} & {2nd} & {3rd} &   \\ \cline{1-4}
		\textbf{Total time of the seq} &  112.4 & 180 & 103.2 &  \\ \hline
		\textbf{ECOHC}~\cite{danelljan2017eco} & \cellcolor{yellow!20} 112.4 & \cellcolor{yellow!20} 145.7 &  \cellcolor{yellow!20} 103.2 &  49.9 \\ \hline
		\textbf{fDSST}~\cite{danelljan2016discriminative} & 70.7 & 78.9 & 103.2 &  186.8 \\ \hline
		\textbf{CACF}~\cite{mueller2017context} & 70.5 & 78.9 & 103.2 &  296.1 \\ \hline
		\textbf{Staple}~\cite{bertinetto2016staple} & 112.4 & 78.9 & 103.2 &  81.9 \\ \hline
		\textbf{KCF}~\cite{henriques2014high} & 70.5 & 78.9 & 103.2 &  \cellcolor{yellow!20} 623.6 \\ \hline
		\hline
	\end{tabular}
	\label{tb:visual_tracking_own}
\end{table}


\subsubsection{Evaluation Results of Visual Preprocessing Module}
\label{sec:Preprocessing_exp}
We evaluated the visual preprocessing module on the collected datasets, with quantitative success rates detailed in~\cref{tb:prepro_snap_tb}. Failures were primarily due to occlusion, fast motion, blur, or fiducials being out of the camera's view.

\begin{figure}
    \centering
    \includegraphics[width=1.0\linewidth]{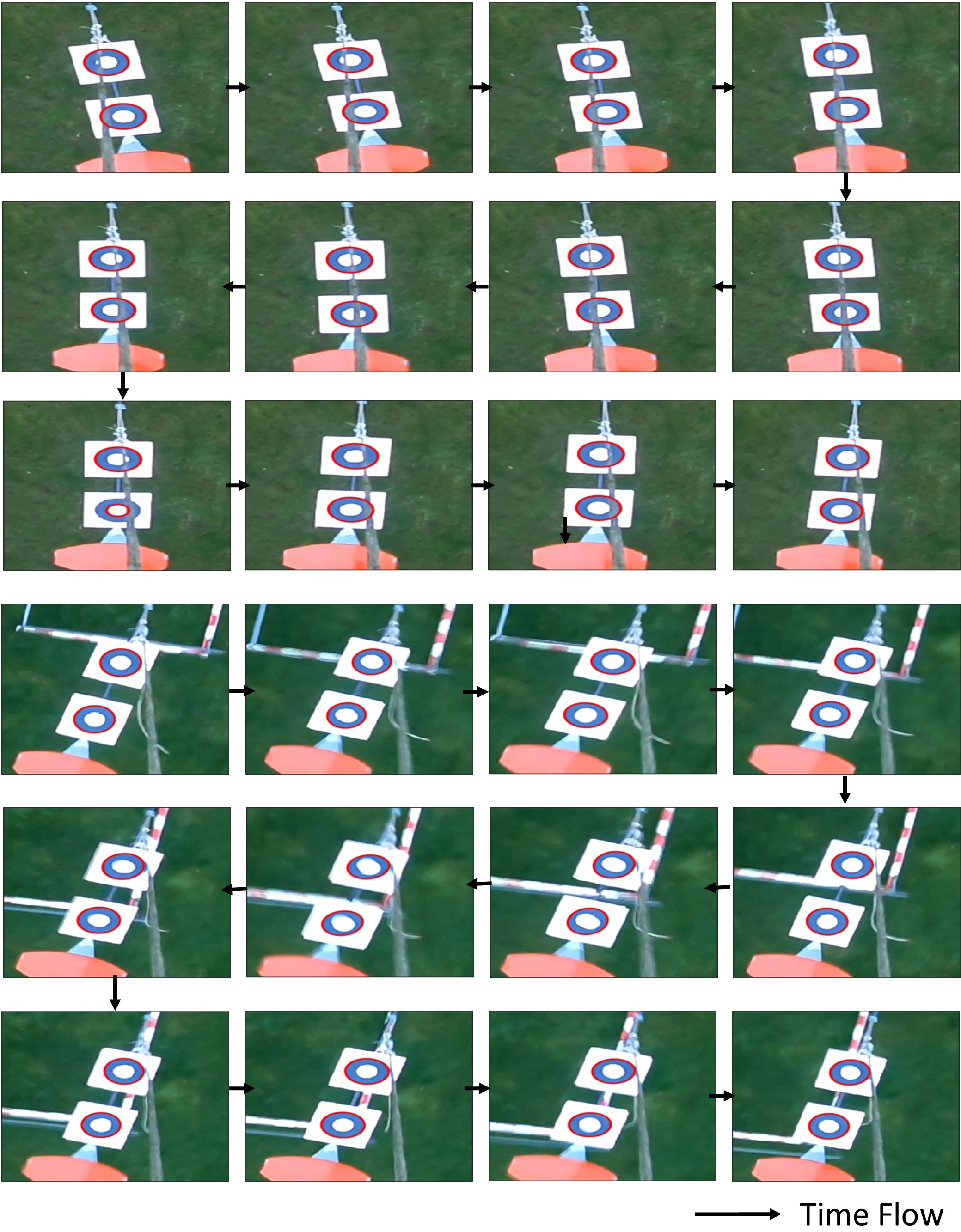}
    \caption[Representative visual results of ellipse detection on collected datasets: partial occlusion.]{Representative visual results of ellipse detection on collected datasets. The black arrow indicates the time flow. The results highlights the performance under partial occlusion.}
    \label{fig:vs_own_1}
\end{figure}

Subsequently, the performance of the ellipse detection module was assessed using the same datasets. Success rates for each dataset are provided in~\cref{tb:prepro_snap_tb}, with representative results shown in~\cref{fig:vs_own_1}. The arc-based ellipse detector demonstrates robust performance, effectively handling rope occlusions and extreme tilt angles. However, a few false detections occur when markers are close together, resulting in incorrect arc grouping. To mitigate these issues, future work will focus on increasing the distance between markers and applying additional positional constraints to prevent incorrect grouping of distant arcs.

\begin{table}[!h]
	\centering
	\caption{{Quantitative evaluation results of the visual preprocessing and Ellipse Detection Module on collected datasets: VP (visual preprocessing), VPED (visual preprocessing \& ellipse detection)}}
	\begin{tabular}{c|c|c|c}
		\hline \hline
		& \multicolumn{3}{c}{\textbf{Success Rate}} \\ \cline{1-4}
        \textbf{Sequence Number} & {1st} & {2nd} & {3rd}   \\ \cline{1-4}
		\textbf{VP Only} & $100\%$ & $98.0\%$ &  $100\%$  \\ \hline
          \textbf{VPED} & $98.8\%$ & $94.4\%$ &  $99.6\%$  \\ \hline
		\hline
	\end{tabular}
	\label{tb:prepro_snap_tb}
\end{table}


\subsection{Sensor Fusion Evaluation Results}
\label{sec:results}
\begin{figure*}[htbp]
    \centering
     \includegraphics[width=0.35\linewidth]{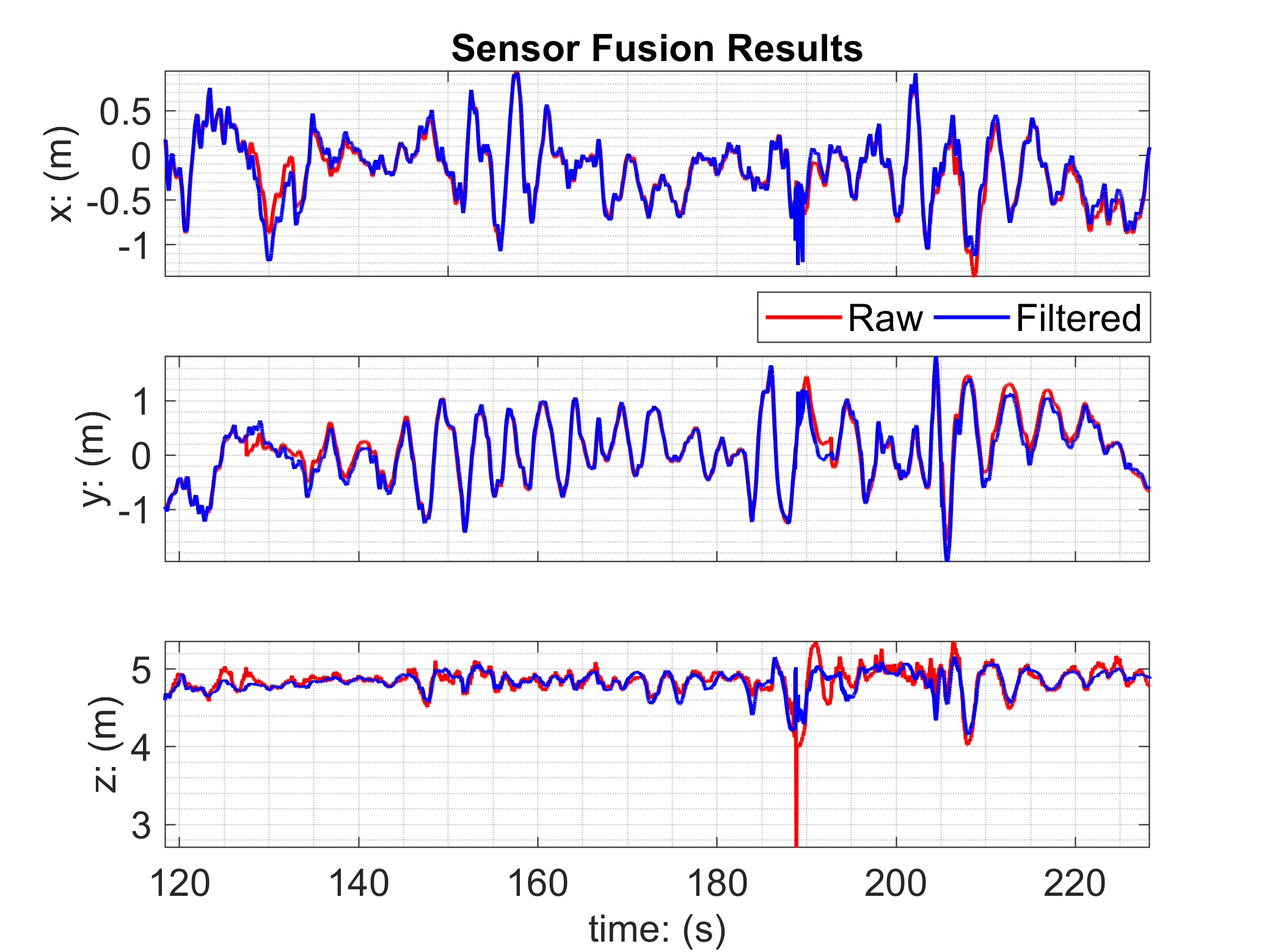}
    \includegraphics[width=0.31\linewidth]{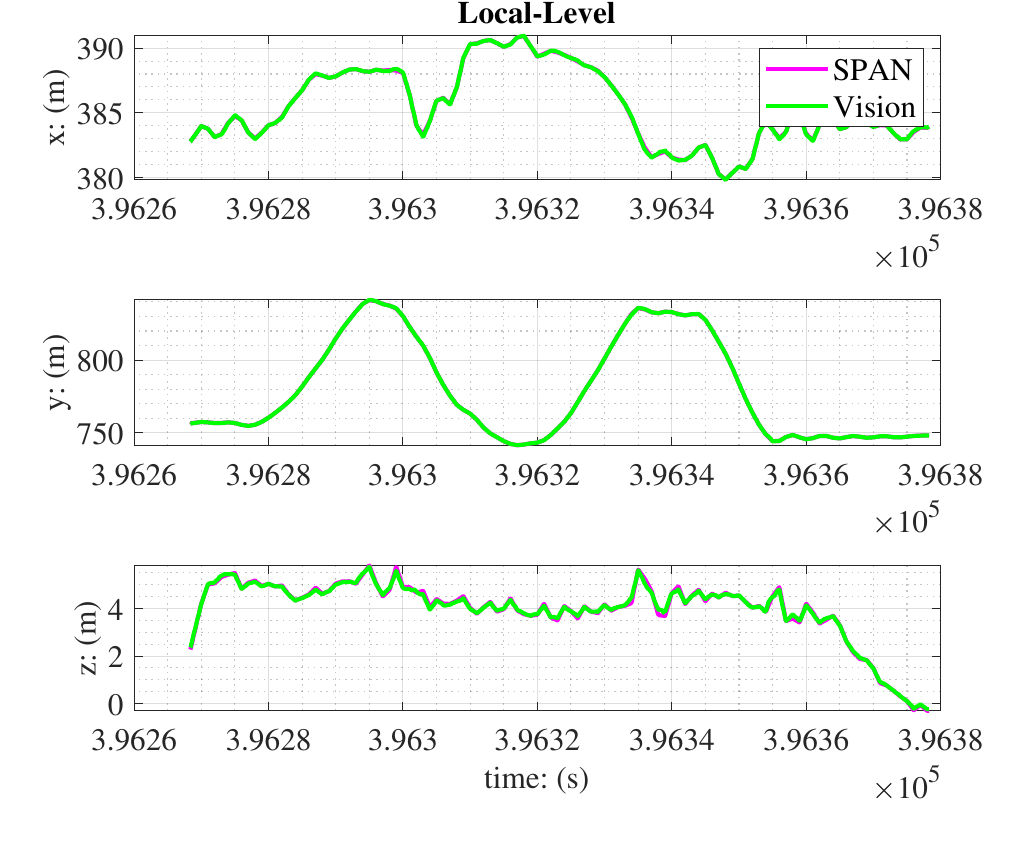}
    \includegraphics[width=0.3\linewidth]{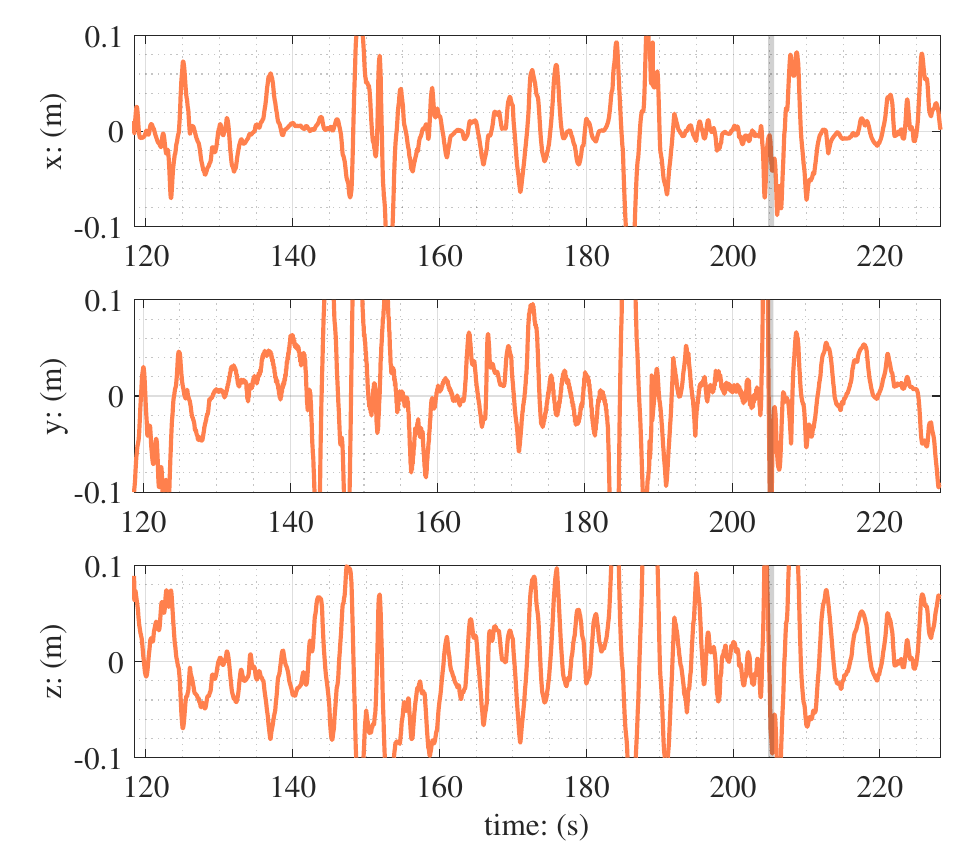}\\
    \centering{\small A. Experiment results: first flight.} \\
    \vspace{0.2em}
    \includegraphics[width=0.35\linewidth]{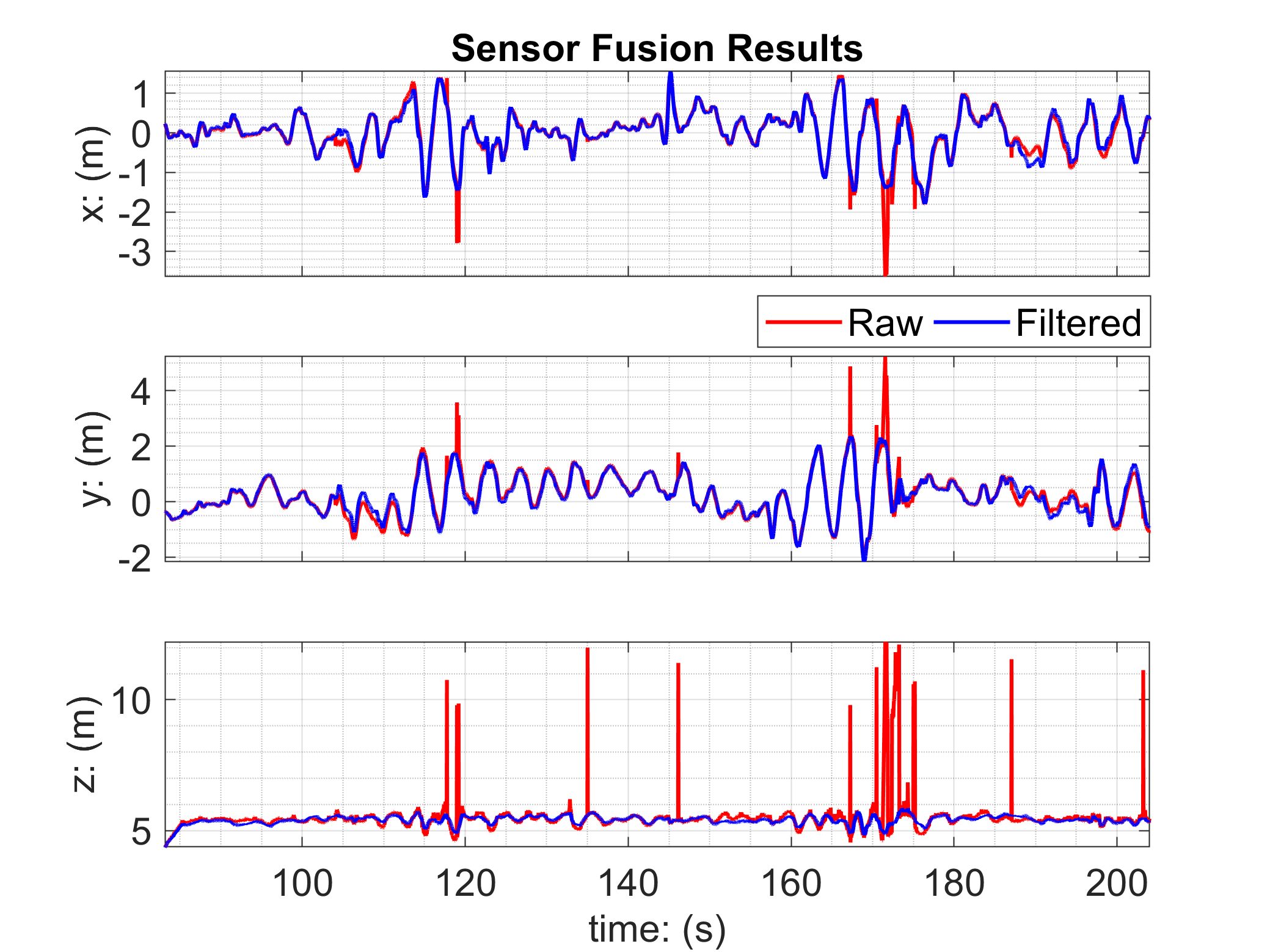}
    \includegraphics[width=0.31\linewidth]{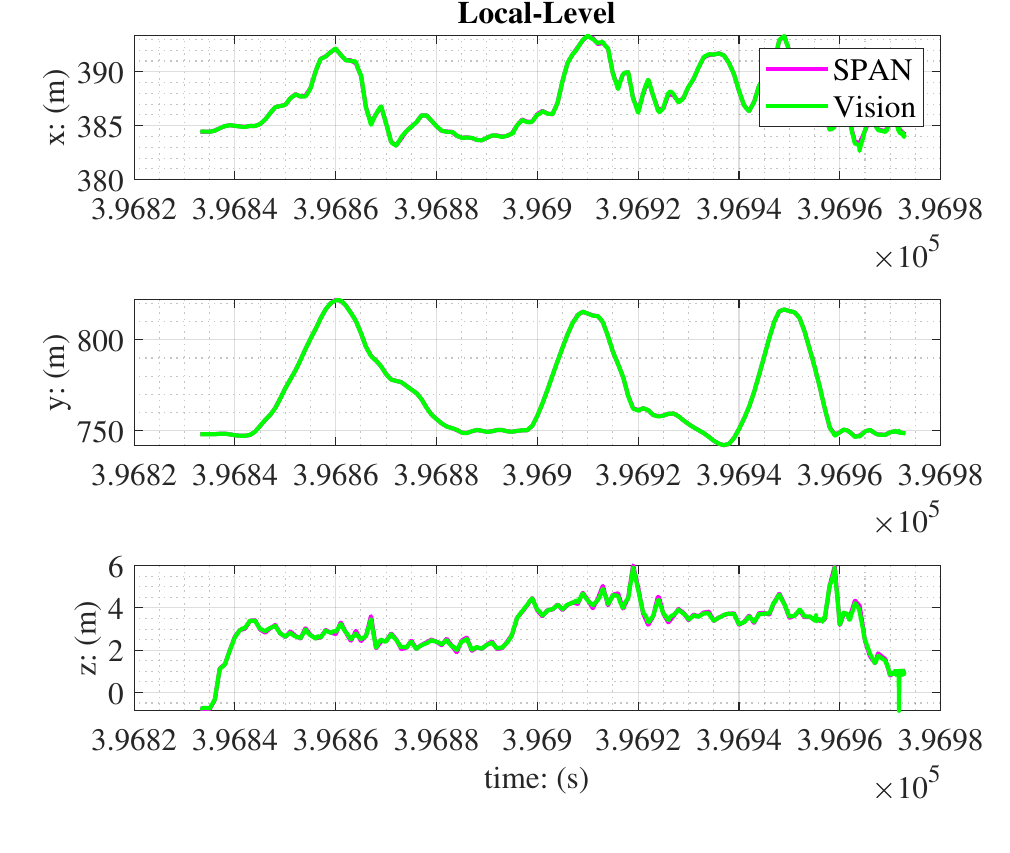}
    \includegraphics[width=0.3\linewidth]{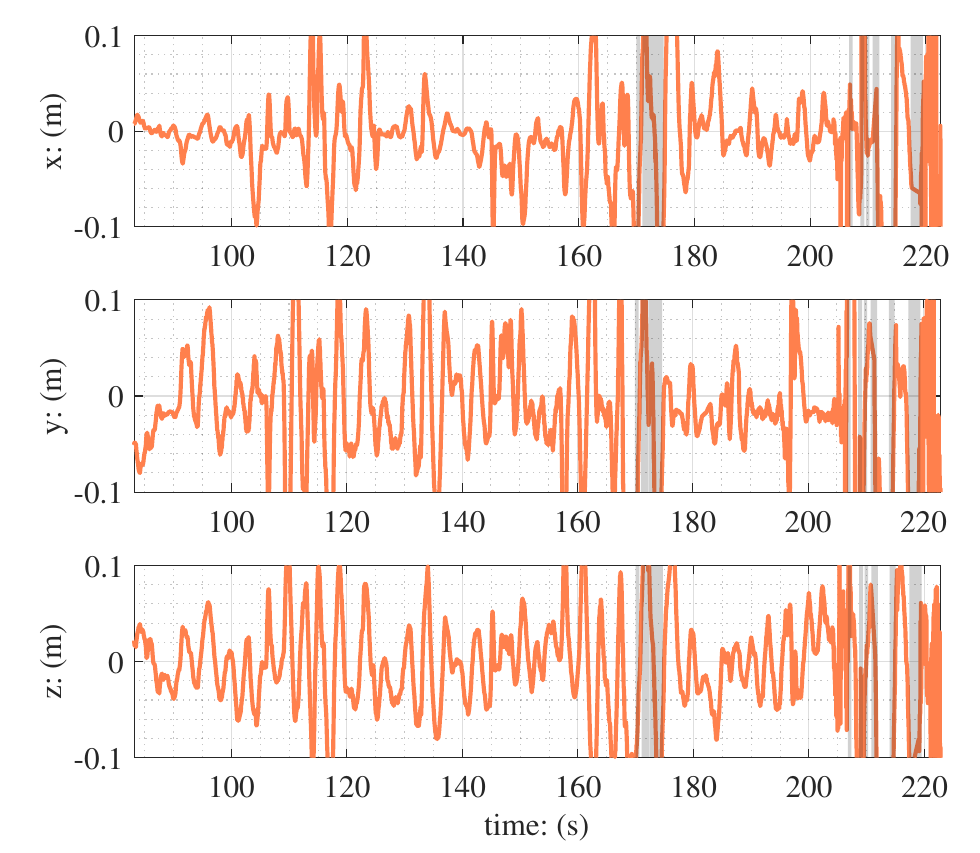}\\
    \centering{\small B. Experiment results: second flight.} \\
    \vspace{0.2em}
     \includegraphics[width=0.35\linewidth]{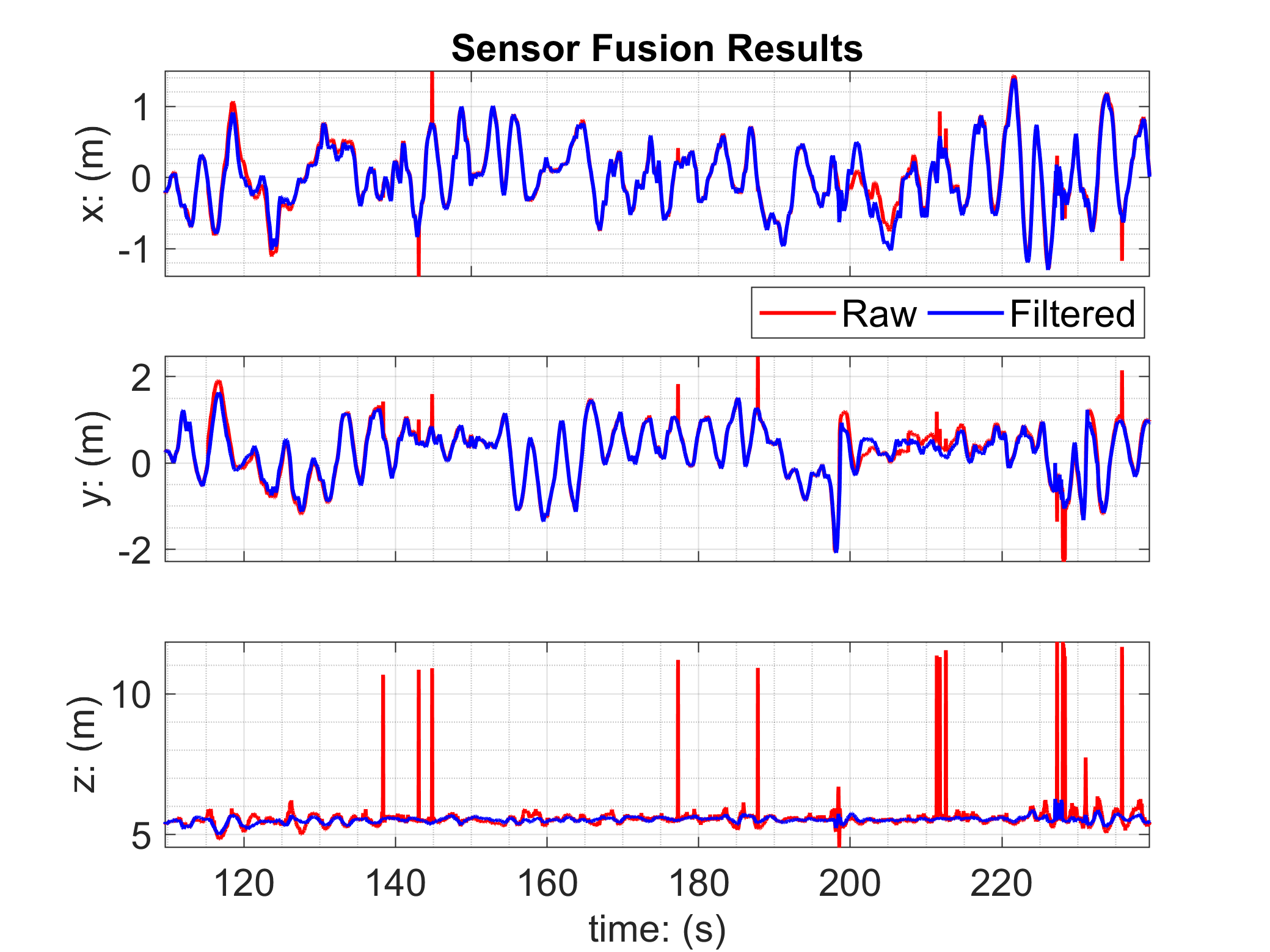}
    \includegraphics[width=0.31\linewidth]{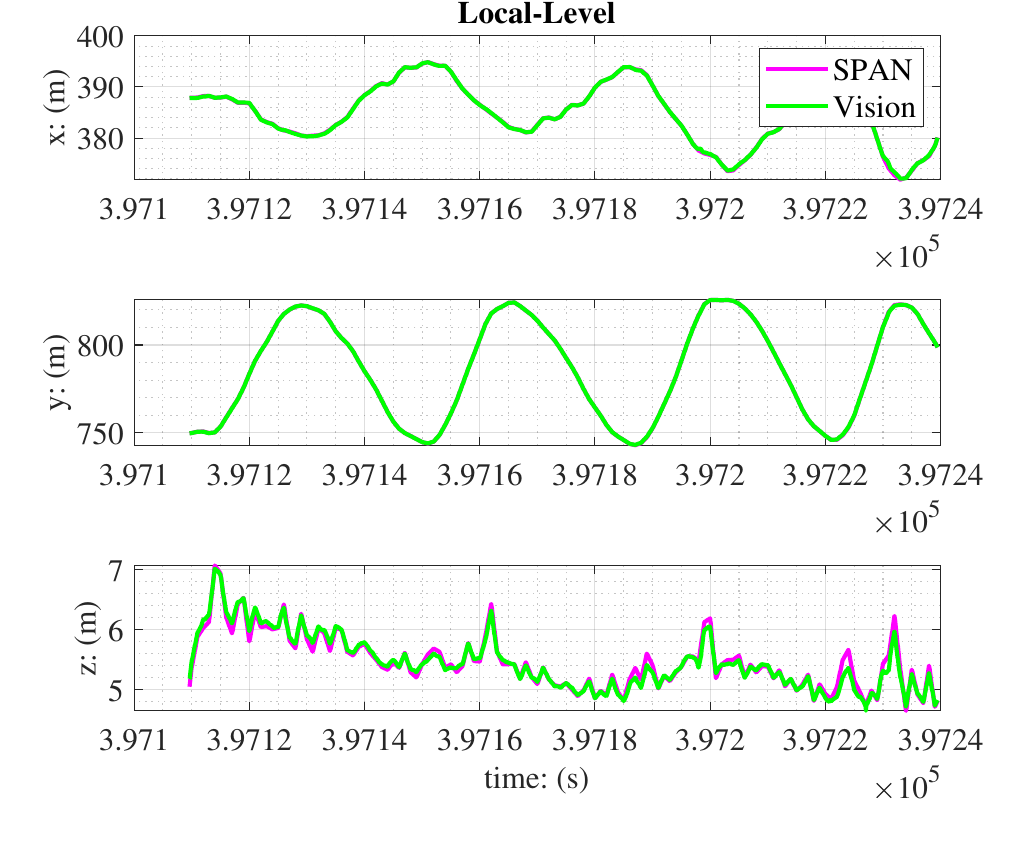}
    \includegraphics[width=0.3\linewidth]{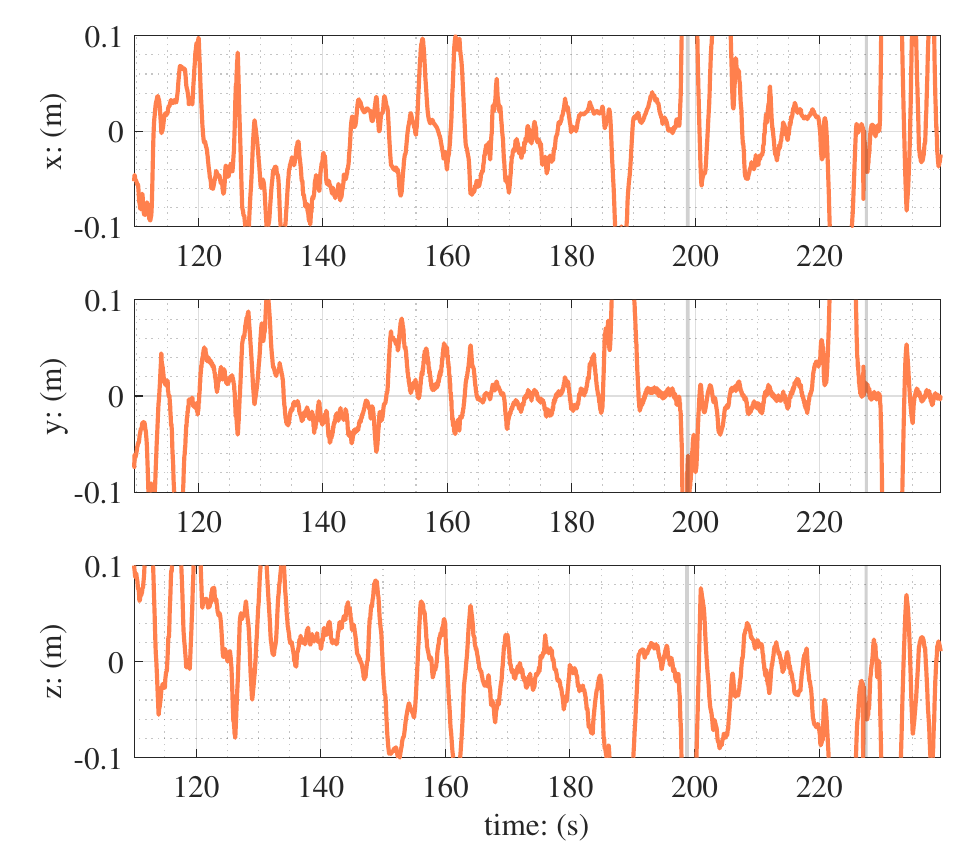}\\
    \centering{\small C. Experiment results: third flight.} \\
    \caption{From left to right: plots of vision measurements and filtered states, positioning results in the local level coordinate, and positioning error plots of the east, north, and up axis, respectively. Visual outputs are not available inside light-dark shaded regions.}    
    \label{fig:exp1_type1}
\end{figure*}

The results of the first flight are shown in Fig.~\ref{fig:exp1_type1} A. The first column displays raw visual (red) measurements and results after sensor fusion (blue). The second column contrasts the positioning results from the SPAN system (magenta path) with those estimated by the proposed approach (light green path), both in the local level coordinate frame. The last column provides error plots in the local level coordinate system, with light-dark shaded regions indicating periods with no visual output due to fiducial detection failures.

In the second flight test, we used larger fiducial markers and more complex flight paths to assess positioning accuracy. Fig.~\ref{fig:exp1_type1} B shows filtered results (left column), positioning results (middle column), and error plots (right column), demonstrating the impact of marker size on accuracy.

The third flight test used the same configuration as the second test but involved more agile UAV maneuvers to evaluate robustness under higher speeds and varying attitudes. Fig.~\ref{fig:exp1_type1} C illustrates the flight paths, filtered positioning results (first column), and error plots (last two columns), showcasing the system's effectiveness in maintaining accurate positioning during agile maneuvers and varying UAV attitudes.

\begin{table}[!h]
	\centering
	\caption[Quantitative Results of the Three Flight Tests: Mean Error and Standard Deviation of East, North, and Up Axes.]{Quantitative Results of the Three Flight Tests: Mean Error and Standard Deviation of East, North, and Up Axes. \textbf{ME} represents Mean Error and \textbf{SD} represents Standard Deviation.}
	\begin{tabular}{|c|c|c|c|c|}
	    \toprule
		\toprule
		\multirow{2}{*}{\textbf{Flight No.}}&\multirow{2}{*}{\textbf{Results}}& \multicolumn{3}{c|}{\textbf{Proposed Approach (m)}} \\ \cline{3-5}
        & & \multicolumn{1}{c}{\textbf{East}}& \multicolumn{1}{c|}{\textbf{North}} & \multicolumn{1}{c|}{\textbf{Up}} \\\cline{1-5}
        \multirow{2}{*}{1} & \textbf{ME} & 0.0332 & 0.0510 & 0.0732 \\\cline{2-5}
        & \textbf{SD} & 0.0391 & 0.0775 & 0.0689 \\\cline{1-5}
        \multirow{2}{*}{2} & \textbf{ME} & 0.0338 & 0.0553 & 0.0412 \\\cline{2-5}
        & \textbf{SD} & 0.0561 & 0.0700 & 0.0409 \\\cline{1-5}
        \multirow{2}{*}{3} & \textbf{ME} & 0.0619 & 0.0423 & 0.0528 \\\cline{2-5}
        & \textbf{SD} & 0.1044 & 0.0709 & 0.0554 \\\cline{1-5}
        \bottomrule 
	\end{tabular}
	\label{tb:exp_res}
\end{table}

\subsection{Overall Performance} 
We show the quantitative results in~\cref{tb:exp_res}. The following qualitative observations has been made from the developed system:
\begin{itemize}
    \item Compared to our previous work~\cite{hu2021toward}, the proposed method demonstrates superior performance with higher vision estimation rates and improved positioning results.  Additionally, the developed filter, utilizing a general motion model driven by the IMU, yields better pose estimation. Further improvements can be achieved by fine-tuning covariance parameters, though this requires precise sensor modeling.
    \item The proposed method achieves centimeter-level accuracy for each axis and maintains a decimeter-level overall 3D position error, demonstrating its precision in positioning outcomes.
    \item Larger fiducial markers could improve accuracy, but constraints on weight and aerodynamics limit their size. Enhancing image resolution might offer a solution, though it increases processing time. Balancing accuracy with practical limitations is essential.
    
\end{itemize}

\section{CONCLUSION}
\label{sec:con}
This paper presents a vision-based cooperative localization system for airborne magnetic surveying. The system incorporates a visual processing module that integrates detection, tracking, and a robust arc-based fiducial detection algorithm, ensuring high success rates and robustness against occlusions and varying viewpoints. It also features an improved manifold-based sensor fusion algorithm that combines local and global positioning results, delivering reliable and accurate positioning measurements. Real flight evaluations show centimeter-level accuracy on individual axes and decimeter-level accuracy in 3D. The system's observability has been validated, proving its effectiveness in real-world applications such as surveying, cargo transport, and aerial manipulation.

\appendices
\section*{APPENDIX}
\subsection{Observability Analysis}
\label{sec:obs_type2}
Considering a general nonlinear system described as follows:
$$
\begin{aligned}
\dot{\mathbf{x}}=\mathbf{f}(\mb{x},\mb{u})\ , \mb{y}=\mb{h}(\mb{x})\ ,
\end{aligned}
$$
where $\mb{x}$, $\mb{u}$, and $\mb{y}$ represent the states, inputs, and outputs, respectively. Let $\mathbf{O}$ being the observability space defined by $\mathbf{O}=[\mathbf{h}^\top\ \  L_\mathbf{f}^1\mathbf{h}(\mb{x})^\top\ \ L_\mathbf{f}^2\mathbf{h}(\mb{x})^\top \ \ \cdots]^\top$ where $L_\mathbf{f}^1\mathbf{h}$ denotes the Lie derivative of vector field $\mathbf{h}$ along vector field $\mathbf{f}$. Then, the system is considered observable given the observability matrix $d\mb{O}=[d\mb{h}^\top\ dL_\mathbf{f}^1\mathbf{h}(\mb{x})^\top\ dL_\mathbf{f}^2\mathbf{h}(\mb{x})^\top\ \cdots]^\top$ is of full rank~\cite{hermann1977nonlinear}. We prove here that the addressed system is observable.

\begin{proof}[Observability Analysis.]
In the lens of the addressed problem, we have to derive the observability matrix as follows. 
First, it is easy to obtain
\begin{equation}
\scalebox{0.89}{$
\begin{aligned}
d\mb{h}&=\left[
\begin{array}{cccccc}
\mb{I} & \mb{0} & \mb{0} & 0 & \mb{0} & \mb{0} \\
\mb{0} & \mb{R}_{\text{G}}^{\text{C}} & \mb{0} & 0 & 
    \mb{R}_{\text{G}}^{\text{C}}\mb{R}_{\text{L}}^{\text{G}} & \mb{I} \\
\mb{0} & \frac{1}{\|\mb{p}^{\text{G}}_\text{L}-\mb{p}^\text{G}_\text{B}\|}(\mb{p}^{\text{G}}_\text{L}-\mb{p}^\text{G}_\text{B})^\top & \mb{0} & -1 & \mb{0} & \mb{0}
\end{array}
\right] \\ &= \left[
\begin{array}{cc}
\mb{I} & \mb{0} \\
\mb{0} & d\mb{h}^{\prime}
\end{array}
\right]\ .
\end{aligned}
$}
\label{eq:v1v2v3}
\end{equation}
It is clear that the orientation state $\mb{R}_\text{L}^\text{G}$ is decoupled with other states and also observable and the left-upper corner block is of full rank. For brevity, we only analyze the observability for the remaining part $d\mb{h}^{\prime}$.

Next, the $L_\mathbf{f}^1\mathbf{h}^\prime(\mb{x})$ is defined as 
\begin{equation}
\scalebox{0.89}{$
L_\mathbf{f}^1\mathbf{h}^\prime(\mb{x})=\left[
\begin{array}{ccccc}
\mb{0} & \mb{R}_{\text{G}}^{\text{C}} & 0 & \mb{0} & \mb{0} \\
 \underbrace{\frac{\boldsymbol{\rho}^\top \mb{v}_{\text{L}}^\text{G}}{\|\boldsymbol{\rho}\|^{3/2}}\boldsymbol{\rho}^\top + \frac{1}{\|\boldsymbol{\rho}\|}{\mb{v}_{\text{L}}^\text{G}}^\top}_{\times_1}  & \underbrace{\frac{\boldsymbol{\rho}^\top}{\|\boldsymbol{\rho}\|}}_{\times_2} & 0 & \mb{0} & \mb{0}
\end{array}
\right]$}\ ,
\end{equation}
where we use $\boldsymbol{\rho}=\mb{p}^{\text{G}}_\text{L}-\mb{p}^\text{G}_\text{B}$ for brevity. 

By following the same routine, we have $L_\mathbf{f}^2\mathbf{h}^\prime(\mb{x})$ as follows:
\begin{equation}
\scalebox{0.9}{$
\begin{aligned}
&L_\mathbf{f}^2\mathbf{h}^\prime(\mb{x})=\\
&\left[
\begin{array}{ccccc}
\mb{0} & \mb{0} & 0 & \mb{0} & \mb{0}\\
\underbrace{\frac{\partial \times_1 \mb{v}_\text{L}^\text{G}}{\partial \mb{p}_\text{L}^\text{G}}+\frac{\partial \times_2(\mb{R}_\text{L}^\text{G}\mb{a}-\mb{g})}{\partial \mb{p}_\text{L}^\text{G}}}_{\times_3} & \underbrace{\frac{\partial \times_1\mb{v}_\text{L}^\text{G}}{\partial \mb{v}_\text{L}^\text{G}}}_{\times_4} & 0 & \mb{0} & \mb{0} \\
\end{array}
\right]\ .
\end{aligned}$}
\end{equation}
Similarly, $L_\mathbf{f}^3\mathbf{h}^\prime(\mb{x})$ can be derived by taking the partial derivatives of the corresponding terms:
\begin{equation}
\scalebox{0.89}{$
\begin{aligned}
&L_\mathbf{f}^3\mathbf{h}^\prime(\mb{x})=\\
&\left[
\begin{array}{ccccc}
\mb{0} & \mb{0} & 0 & \mb{0} & \mb{0}\\
\underbrace{\frac{\partial \times_3 \mb{v}_\text{L}^\text{G}}{\partial \mb{p}_\text{L}^\text{G}}+\frac{\partial \times_4(\mb{R}_\text{L}^\text{G}\mb{a}-\mb{g})}{\partial \mb{p}_\text{L}^\text{G}}}_{\times_5} & \underbrace{\frac{\partial \times_3\mb{v}_\text{L}^\text{G}}{\partial \mb{v}_\text{L}^\text{G}}}_{\times_6} &0 &\mb{0} & \mb{0} \\
\end{array}
\right]\ .
\end{aligned}
$}
\end{equation}

Luckily, the above Lie derivatives are sufficient to evaluate the observability so that we do not need to derive more Lie derivatives. 
Apparently, the partial observability matrix can be written as
\begin{equation}
\scalebox{0.94}{$
\begin{aligned}
d\mb{O}_4&=\left[
\begin{array}{c}
    d\mb{h}^\prime  \\
    dL_\mb{f}^1\mb{h}^\prime \\
     dL_\mb{f}^2\mb{h}^\prime \\
     dL_\mb{f}^3\mb{h}^\prime \\
\end{array}
\right]=\left[
\begin{array}{ccccc}
    \mb{R}_{\text{G}}^{\text{C}} & \mb{0} & 0 & \mb{R}_{\text{L}}^{\text{C}} & \mb{I}  \\
    \frac{\boldsymbol{\rho}}{\|\boldsymbol{\rho}\|} & \mb{0}  & -1 & \mb{0} & \mb{0} \\
    \mb{0} & \mb{R}_{\text{G}}^{\text{C}} & 0 & \mb{0} & \mb{0} \\
    \times_1 & \times_2 & 0 & \mb{0} & \mb{0} \\
    \mb{0} & \mb{0} & 0 & \mb{0} & \mb{0} \\
    \times_3 & \times_4 & 0 & \mb{0} & \mb{0} \\
    \mb{0} & \mb{0} & 0 & \mb{0} & \mb{0} \\
    \times_5 & \times_6 & 0 & \mb{0} & \mb{0} \\
\end{array}
\right]\ ,
\end{aligned}
$}
\end{equation}
where each $\times_k$ denotes a nonzero element. 
Since the matrix $d\mb{O}_4$ has more rows than columns, we analyze its column space. 
If $\mb{R}_\text{L}^\text{C}=\mb{R}_\text{G}^\text{C}\mb{R}_\text{L}^\text{G}$ is not an identity matrix, then we need at least seven basis vectors are needed for the last three states $l,\ \Delta \mb{t}^\text{C}_\text{B},\  \mb{t}_\text{F}^\text{L}$, i.e. $\mathbf{e}_i, i=7,\cdots,13, \mathbf{e}_i=[0,\cdots,\underset{i}{1},\cdots,0]$. Then, we take the sub-matrix $\left[\begin{array}{cc} \mb{0} & \mb{R}_{\text{G}}^{\text{C}} \\ \times_1 & \times_2 \end{array}\right]$ and evaluate its determinant. It is clear that if $\mb{v}_\text{L}^\text{G}$ and $\boldsymbol{\rho}$ are not zero, the determinant is nonzero so that the rank of the submatrix is 2. To violate the condition, the load has to be always static ($\mb{v}_\text{L}^\text{G}=\mb{0}$) or the load is exactly attached to the body mass center ($\boldsymbol{\rho}=0$), which is uncommon in practice. Overall speaking, the partial observability matrix is full rank except in some rare cases. Therefore, we consider the system is observable, which concludes the proof.
\end{proof}





\bibliographystyle{IEEEtaes}
\bibliography{mybibfile,det_track}

\end{document}